\documentstyle[prd,aps,psfig]{revtex}

\def\dateornot{\date{Accepted for pubmication in Physical Review D}}
\def\teq#1{$\, #1\,$}                           % text equation
\def\dover#1#2{\hbox{${{\displaystyle#1 \vphantom{(} }\over{
   \displaystyle #2 \vphantom{(} }}$}}

\def\erg{\varepsilon}
\def\eperp{e_{\perp}}
\def\epar{e_{\parallel}}
\def\vsp{\vphantom{\Bigl\{}}

\def\vspm{\vphantom{\biggl\{}}
\def\vspl{\vphantom{\Biggl\{}}
\def\fsc{\alpha_{\rm f}}
\def\pll{\perp\to\parallel\parallel}
\def\lpl{\parallel\to\perp\parallel}
\def\ppp{\perp\to\perp\perp}
\def\pmb#1{\setbox0=\hbox{#1}%
  \kern-0.0125em\copy0\kern-\wd0
  \kern0.025em\copy0\kern-\wd0
  \kern-0.0125em\raise0.0433em\box0 }
             \font\sevenrm=cmr7

\begin{document}
\draft
%
%\tightornot
%
\title{MAGNETIC PHOTON SPLITTING: THE S-MATRIX\\
        FORMULATION IN THE LANDAU REPRESENTATION}
\author{
Matthew G. Baring\thanks{Universities Space Research
    Association. Email: baring@lheavx.gsfc.nasa.gov}    }
\address{Laboratory for High Energy Astrophysics, Code 661,\\
   NASA Goddard Space Flight Center, Greenbelt, MD 20771, USA}
\dateornot
\maketitle

\begin{abstract}
Calculations of reaction rates for the third-order QED process of
photon splitting \teq{\gamma\to\gamma\gamma} in strong magnetic fields
traditionally have employed either the effective Lagrangian method or
variants of Schwinger's proper-time technique.  Recently, Mentzel, Berg
and Wunner \cite{mbw94} presented an alternative derivation via an
S-matrix formulation in the Landau representation.  Advantages of such
a formulation include the ability to compute rates near pair resonances
above pair threshold.  This paper presents new developments of the
Landau representation formalism as applied to photon splitting,
providing significant advances beyond the work of \cite{mbw94} by
summing over the spin quantum numbers of the electron propagators, and
analytically integrating over the component of momentum of the
intermediate states that is parallel to field.  The ensuing tractable
expressions for the scattering amplitudes are satisfyingly compact, and
of an appearance familiar to S-matrix theory applications.  Such
developments can facilitate numerical computations of splitting
considerably both below and above pair threshold.  Specializations to
two regimes of interest are obtained, namely the limit of highly
supercritical fields and the domain where photon energies are far
inferior to that for the threshold of single-photon pair creation.  In
particular, for the first time the low-frequency amplitudes are simply
expressed in terms of the Gamma function, its integral and its
derivatives.  In addition, the equivalence of the asymptotic forms in
these two domains to extant results from effective
Lagrangian/proper-time formulations is demonstrated.
\end{abstract}

\section{INTRODUCTION}
\label{sec:intro}

The third-order quantum electrodynamical process of photon splitting
\teq{\gamma\to\gamma\gamma} in a strong magnetic field, currently
popular in several astrophysical models of different neutron star
sources, was first studied over three decades ago.  Due to analytic
complexities encountered when investigating this interaction, it was
not until the beginning of the 1970s that a body of correct and
uncontroversial results emerged.  These early splitting calculations
used either effective Lagrangian \cite{Adl70,Bial70,Adler71} or
variations of Schwinger's proper-time techniques
\cite{PR72,Stone79,bms86}, the expediency of which yielded compact
analytic forms for the rates \teq{R} when specializing to low energy
(\teq{R\propto \omega^5}) or low field (\teq{R\propto B^6}) cases.
After a hiatus of nearly two decades, photon splitting became of
interest again in the literature \cite{bms96,as96,bh97apj,wbm98}
following the publication of an S-matrix calculation in the Landau
representation of its rates by Mentzel, Berg and Wunner\cite{mbw94},
specifically because of their contention that the earlier works cited
above had seriously underestimated the strength of this process.  The
rates computed in \cite{mbw94} were later retracted in \cite{ww97},
with a sign error in their numerical coding having been discovered and
corrected.  Mentzel et al.'s analytic derivation was the first
comprehensive presentation of the application of a Landau
representation technique specifically to magnetic photon splitting,
though the QED formalism presented by Melrose and Parle
\cite{mpII,mpIII} virtually provided an equivalent enunciation of such
S-matrix forms for splitting amplitudes.  More recently, Weise, Baring
\& Melrose \cite{wbm98} confirmed the analytic derivation of
\cite{mbw94}.  The Landau representation calculations and most of the
earlier effective Lagrangian and proper-time presentations were
generally applicable to non-dispersive regimes below the pair creation
threshold (\teq{\hbar\omega =2mc^2}), where the momentum vectors of the
initial and final photons are collinear, and arbitrary field
strengths.

Below pair threshold, the effective Lagrangian approach of
\cite{Adl70,Bial70,Adler71} and the proper-time calculations in
\cite{PR72,Stone79,bms86,bms96} appear much more amenable for the
purposes of numerical evaluation than the S-matrix formulation in the
Landau representation.  This arises because effective Lagrangian and
proper-time (collectively referred to by the label ELP here) methods
produce results that involve triple integrals over relatively simple
(hyperbolic and exponential) functions, while the S-matrix amplitudes
integrate over the parallel momentum \teq{p_z} and include a triple
summation over the Landau level quantum numbers of the intermediate
pair states.  Both techniques start from different but equivalent
\cite{mpI} forms of the electron propagator, and hence S-matrix
computations \cite{mbw94,wbm98} should yield identical results to
proper-time numerics
\cite{Adl70,Bial70,Adler71,PR72,Stone79,bms86,bms96}.  For the specific
case of magnetic pair creation \teq{\gamma\to e^+e^-}, such an
equivalence of the S-matrix and proper-time methods has been
demonstrated\cite{te74,dh83}, but only via continuous asymptotic
approximations that smoothly average out the exact ``sawtooth''
resonance structure.  Yet the S-matrix Landau representation approach
explicitly retains the resonances in the scattering amplitudes above
pair threshold, whereas the ELP methods eliminate such information
early during developments.  Photon splitting becomes effectively
first-order in \teq{\fsc} at any one of a multitude of pair resonances,
generated when the intermediate states become ``on-shell.''   Hence it is
quite possible that splitting can compete effectively with pair
creation as a photon absorption mechanism above pair threshold.
Ascertaining whether this is true is an interesting physics question.
Moreover, if splitting is approximately as probable as pair creation
above threshold, then it manifestly changes the character of vacuum
dispersion, so that quadratic (and by inference perhaps higher order)
contributions to the vacuum polarization tensor become significant
relative to the standard linear ones used in the derivation
\cite{Adler71} of kinematic selection rules for splitting.  Hence, the
generation of exact and compact expressions for the rates for
\teq{\gamma\to\gamma\gamma} valid both below and above pair creation
threshold is clearly a worthwhile enterprise from a physics
perspective.

Developed expressions for the rates for photon splitting are also
important for astrophysical applications of this process, particularly
to effect efficient and accurate computations of such rates.  These
applications have so far focused on neutron star magnetospheres,
primarily on models of soft gamma repeaters (SGRs) and
strongly-magnetized pulsars, both being extremely topical in the
astrophysics community at present.  The potential importance of
splitting in neutron star environments was suggested by
\cite{Adler71,mitrof86,b88}.  Possible formation of splitting cascades
has been explored in models of SGR transient outbursts as a means of
softening the spectrum efficiently with no production of pairs
\cite{b95apjl,bh95apss,hb96aip,ccfh96aip}.  If both polarizations can
split, or if polarization switching is active during SGR outbursts,
then the properties of the splitting cross-section guarantee emergent
spectra in the observed range (20--150 keV) and of the observed shape
for all fields in excess of around \teq{10^{14}}Gauss
\cite{b95apjl,bh95apss,hb96aip}, provided that the emission region is
not concentrated near the polar cap.  The spectral properties of SGRs
in quiescent emission appear to be distinct from those during
outburst.  Pulsations and temporal increases of their periods (i.e.
spin-down) have now been observed \cite{kouv99nat,hurl99nat,kouv99apjl}
for two of the four confirmed SGRs (SGR 1806-20 and SGR 1900+14),
leading to inferences of fields in the vicinity of \teq{10^{15}}Gauss.
The connection between these pulsars of extremely high magnetization,
so-called {\it magnetars}, and conventional radio/X-ray/gamma-ray
pulsars is not well-understood.  Baring \& Harding \cite{bh98apjl}
postulated that radio quiescence, a property of the SGRs, may be common
in magnetars due to the efficient action of photon splitting and other
effects in suppressing the creation of pairs.  Photon splitting also
has spectral implications for such pulsars with more modest fields:
\cite{hbg97apj} demonstrated that the unusual absence of \teq{>30}MeV
emission in the gamma-ray pulsar PSR 1509-58 (whose spin-down field is
\teq{\sim 3\times 10^{13}}Gauss) can naturally be explained by the
operation of \teq{\gamma\to\gamma\gamma} in the intense magnetic and
gravitational fields near its surface.

Several desirable goals are immediately identifiable on the basis of
this historical path for the study of the physics of photon splitting,
and the needs of the astrophysics community.  It would be satisfying
(i) to obtain analytic expressions for rates that are valid above pair
creation threshold using the Landau representation methodology, (ii) to
know whether the analytic formalism of Mentzel et al. \cite{mbw94} can
be developed and simplified, and (iii) to demonstrate a formal
equivalence between this S-matrix Landau representation approach and
extant results from proper-time/effective Lagrangian techniques.  This
paper addresses these issues, using the verified analytic formalism of
Mentzel et al.  as the starting point for mathematical developments.
The analysis here considers all the polarization modes that are
permitted by the CP invariance symmetry (\teq{\pll}, \teq{\ppp} and
\teq{\lpl}), and applies for collinear momenta of the incoming and
outgoing photons, i.e. when the effects of vacuum dispersion are
neglected.  A significant development provided in this paper is the
dramatic simplification incurred by algebraically performing the
summation over the spin states that are incorporated in the electron
propagators.  The resulting expressions in Section~\ref{sec:reduce}
(first stated in \cite{wbm98}) are relatively compact, and of an
appearance familiar to Landau representation/S-matrix theory
applications to magnetized environments (i.e. including associated
Laguerre functions).  Furthermore, here the integrations over the
momentum parallel to the field are performed analytically for the first
time in Section~\ref{sec:pzint}, rendering the splitting rates in most
amenable forms (see Eqs.~[\ref{eq:totratefin}] and [\ref{eq:Mform}])
that are optimal for numerical applications: the analytic forms
presented consist of just triple summations over Landau level quantum
numbers of the intermediate states.  These general results are valid
both below and above pair threshold at non-resonant photon energies,
and provide substantial advances over the work of \cite{mbw94}; they
are much more suitable for numerical evaluation since many
cancellations have been eliminated algebraically.

Two specializations are discussed in Section~\ref{sec:limits},
primarily to (partially) demonstrate equivalence of the Landau
representation formalism presented here with extant
proper-time/effective Lagrangian limiting forms for splitting rates,
and simultaneously to serve as a check on the mathematical
manipulations of this paper.  Results are presented for all three
polarization modes permitted by CP invariance in the limit of zero
dispersion.  The first asymptotic regime is (see
Section~\ref{sec:Bgg1}) for highly supercritical fields, \teq{B\gg
B_{\rm c}=m^2c^3/e\hbar}, where in the case of \teq{\pll}, the limit
was found to concur with a recent analytic result that was obtained by
Baier et al.  \cite{bms96}, while new results were obtained for the
other two modes.  In the second specialization, in
Section~\ref{sec:omegall1}, asymptotic results for energies
\teq{\omega\ll mc^2} well below pair creation threshold were obtained,
reproducing the cubic energy dependence of the amplitudes obtained by
other QED techniques.  Moreover, new and compact expressions for the
scattering amplitudes in this low energy limit are derived in terms of
the logarithm of the Gamma function, its integral and their
derivatives.  These simplified forms in
Eqs.~(\ref{eq:Mperptoparparwll1}) and~(\ref{eq:Mperptoperpperpwll1})
are also produced from extant integral forms for splitting matrix
elements derived first in \cite{Adl70,Adler71}, thereby facilitating
the first analytic demonstration of the equivalence of splitting rates
obtained by the S-matrix formulation in the Landau representation and
those derived using Schwinger-type techniques.

\section{THE GENERAL S-MATRIX FORMALISM}
\label{sec:formalism}

The rates for photon splitting within an S-matrix formulation can be
developed using a variety of conventions; here the Landau
representation used by Mentzel, Berg and Wunner \cite{mbw94} is
adopted, and formal developments lead to an independent confirmation of
their analytic derivation.  Specifically, for a field {\bf
B}\teq{=(0,0,B)}, this approach uses a representation of the
electron/positron wavefunctions as eigenstates of the magnetic moment
(or spin) operator \teq{\mu_z} (with \teq{\pmb{$\mu$}=m\pmb{$\sigma$} +
\gamma_5\beta \pmb{$\sigma$} \times [\hbox{\bf p}+e\hbox{\bf
A}(\hbox{\bf x}) ] }) in Cartesian coordinates within the confines of
the Landau gauge \teq{{\bf A}({\bf x})=(0,\, Bx,\, 0)}.  Such states
turn out to be very convenient because they generate useful symmetry
properties; they were identified by Sokolov and Ternov \cite{ST68}, who
dubbed them states of ``transverse polarization.''  Let \teq{e}, \teq{e'}
and \teq{e''} denote the polarizations of the initial and final
(primed) photons (\teq{e,\, e',\, e''\, =\perp,\,\parallel}), and
\teq{k_{\mu}=(\omega, \hbox{\bf k})}, \teq{k'_{\mu}=(\omega', \hbox{\bf
k}')} and \teq{k''_{\mu}=(\omega'', \hbox{\bf k}'')} denote the
absorbed and produced photon four momenta.  Then the total rate for
splitting via the polarization mode \teq{e\to e'e''} can be written,
using Eqs.~(27)--(29) of \cite{mbw94}, in terms of the S-matrix element
\teq{S_{fi}^{(3)}}, which is the sum of six terms \teq{S_{fi,j}^{(3)}}
corresponding to the six viable time-ordering possibilities:
\begin{equation}
   R_{e\to e'e''}\; =\; \dover{1}{2}\, \biggl( \dover{V}{8\pi^3} \biggr)^2
   \,\dover{mc^2}{\hbar} \int d^3k'\, d^3k''\,\dover{1}{T}\,
   \biggl\vert \sum_{j=1,6} S_{fi,j}^{(3)} \biggr\vert^2\quad ,
 \label{eq:totrate}
\end{equation}
where \teq{V} and \teq{T} denote the volume and time associated with
the interaction calculation and the factor of \teq{1/2} out the front
avoids double counting of the final states.  The priming convention
adopted throughout the paper is one and two primes for the produced
photons and no prime for the initial photon.  Since the S-matrix
element contains a delta function
\teq{\delta^4(k_{\mu}-k'_{\mu}-k''_{\mu})} prescribing four-momentum
conservation for splitting, it is squared in the usual way using
\teq{\vert\delta^4(k_{\mu}-k'_{\mu}-k''_{\mu}) \vert^2\to [VT/(2\pi
)^4]\,\delta^4(k_{\mu}-k'_{\mu}-k''_{\mu})}.  Note that the S-matrix
element should possess a cubic dependence on photon energies when well
below pair creation threshold, due to parity symmetry, photon gauge
invariance, and the antisymmetric nature of the electromagnetic field
tensor; details are discussed in \cite{Adler71}.

Before writing down expressions for the S-matrix element terms, it is
appropriate to identify the dimensionless convention that shall be
adopted throughout this paper.  Since the electron rest mass \teq{m} is
the only mass that enters into this QED problem, we opt to scale all
energies by \teq{mc^2} and momenta by \teq{mc} unless otherwise
specified.  This includes a scaling of \teq{mc^2/\hbar} for photon
frequencies \teq{\omega}.  In the spirit of this convention, we choose
to use the symbol \teq{\erg} to represent dimensionless electron
energies and reserve \teq{E} (\teq{=\erg mc^2}) to denote
``dimensional'' energies as in \cite{mbw94}.  In addition, the magnetic
field will be expressed in terms of the quantum critical field
\teq{B_{\rm c}=m^2c^3/e\hbar} hereafter, so that \teq{B=1} denotes a
field of \teq{4.413\times 10^{13}}Gauss.

The convention for polarizations is identical to that assumed in
\cite{mbw94}, who opted for real polarization vectors with zero time
components.  The polarization states \teq{\perp} and \teq{\parallel}
are defined according to whether the photon's electric vector lies
either perpendicular or parallel (respectively) to the plane containing
the photon's momentum \teq{{\bf k}} and the (uniform) magnetic field
{\bf B} vectors, the convention of
\cite{mbw94,mpII,Stone79,te74,bh97apj,hrw82}.  In the limit of zero
dispersion, three polarization modes are permitted by charge/parity
(CP) invariance in QED, namely \teq{\pll}, \teq{\ppp} and \teq{\lpl}.
However, Adler \cite{Adler71} showed (see also \cite{Shabad75}) that
for weak vacuum dispersion (roughly delineated by \teq{B \lesssim 1}),
where the refractive indices for the polarization states are very close
to unity, energy and momentum could simultaneously be conserved only
for the splitting mode \teq{\pll}.  This kinematic selection rule
applies to gamma-ray pulsar magnetospheres where plasma dispersion is
negligible.  In magnetar models of soft gamma repeaters, where
supercritical fields are employed, strong vacuum dispersion arises.  In
such a regime, it is not clear whether Adler's selection rules still
endure, since his linear dispersion analysis omits higher order
(quadratic) contributions \cite{mpII,mpIII} to the vacuum polarization
tensor (e.g. those that couple to photon absorption via splitting) that
may become significant in supercritical fields.  Furthermore, plasma
dispersion effects, which can nullify the vacuum selection rules, may
be quite pertinent \cite{bm97} to soft gamma repeater magnetospheres,
rendering them distinctly different from those of conventional
pulsars.  Therefore, in the interests of generality, consideration of
all three CP-permitted splitting modes is adopted throughout this
paper.

The derivation of the S-matrix element proceeds along lines identical
to those in Mentzel, Berg \& Wunner \cite{mbw94}, with the result being
an exact reproduction of their analytic formalism, as reported in
Weise, Baring \& Melrose \cite{wbm98}; for details, one is referred
to \cite{mbw94}.  Of the six \teq{S_{fi,j}^{(3)}} contributions to 
Eq.~(\ref{eq:totrate}), it is sufficient to explicitly present just one:
\begin{eqnarray}
  S^{(3)}_{fi,1} & =& -i\, \dover{\pi^2}{16}
   \dover{(4\pi\fsc )^{3/2}B}{\sqrt{\omega\,\omega'\omega''}}\,
   \dover{1}{(2V)^{3/2}}\, \delta^{(4)}(k_{\mu}-k'_{\mu}-k''_{\mu})\;
   \sum_{n\, n'n''} \; \sum_{\sigma\,\sigma'\sigma''}
   \dover{1}{\erg_0\erg_0'\erg_0''}\int dp_z \nonumber\\
  & \times & \dover{ \vsp
   {\cal D}^{++}_{n'n}(\hbox{\bf k}'')
   {\cal D}^{+-}_{n\, n''}(\hbox{\bf k}')
   {\cal D}^{-+}_{n''n'}(\hbox{\bf k}) -
   {\cal D}^{--}_{n\, n'}(\hbox{\bf k}'')
   {\cal D}^{+-}_{n''n}(\hbox{\bf k}')
   {\cal D}^{-+}_{n'n''}(\hbox{\bf k}) }{ \vsp \erg\,\erg'\erg''\;
   (\erg +\erg''+\omega'-i\epsilon )\, (\erg'+\erg''+\omega -i\epsilon ')}
   \,\Biggl\vert_{p'_z=p_z-k_z'',\, p''_z=-p_z-k'_z}\; ,
 \label{eq:mbweq25}
\end{eqnarray}
where \teq{\fsc = e^2/(\hbar c)} is the fine structure constant, and
the energies \teq{\erg} and \teq{\erg_0} are defined in
Eq.~(\ref{eq:erg}) below, with similar definitions for the primed
energies involving primed quantum numbers and momenta of the virtual
electrons.  Using the \teq{J} notation in Eq.~(\ref{eq:Jeval}) below,
\begin{eqnarray}
   {\cal D}^{++}_{n'n}(\hbox{\bf k}'') & = & \vsp
     J(-k''_x \,\vert\, n'-1,\, -k''_y \,\vert\, n,\, 0)\,
     [\kappa'^{\ast}_1\kappa_4 + \kappa'^{\ast}_3\kappa_2]\; e''_-\nonumber\\
   && +\; J(-k''_x \,\vert\, n',\, -k''_y\,\vert\, n-1,\, 0 )\, \vsp
     [\kappa'^{\ast}_4\kappa_1 + \kappa'^{\ast}_2\kappa_3]\; e''_+\nonumber\\
   && -\; \Bigl\{ J(-k''_x \,\vert\, n',\, -k''_y \,\vert\, n,\, 0)\,
      [\kappa'^{\ast}_2\kappa_4 + \kappa'^{\ast}_4\kappa_2] 
     - J(-k''_x \,\vert\, n'-1,\, -k''_y \,\vert\, n-1,\, 0)\,
      [\kappa'^{\ast}_1\kappa_3 + \kappa'^{\ast}_3\kappa_1]\Bigr\} e''_z
      \; ,\nonumber\\
   {\cal D}^{+-}_{n\, n''}(\hbox{\bf k}') & = & \vsp
      J(-k'_x \,\vert\, n-1,\, 0 \,\vert\, n'',\, k'_y)\, [\kappa^{\ast}_1
      \kappa''^{\ast}_2 + \kappa^{\ast}_3\kappa''^{\ast}_4]\; e'_-\nonumber\\
   && +\; J(-k'_x \,\vert\, n,\, 0 \,\vert\, n''-1,\, k'_y)\, \vsp
      [\kappa^{\ast}_2\kappa''^{\ast}_1 + \kappa^{\ast}_4\kappa''^{\ast}_3]
      \; e'_+ \label{eq:calD}\\
   && -\; \Bigl\{ J(-k'_x \,\vert\, n,\, 0 \,\vert\, n'',\, k'_y)\,
      [\kappa^{\ast}_2\kappa''^{\ast}_2 + \kappa^{\ast}_4\kappa''^{\ast}_4]
          - J(-k'_x \,\vert\, n-1,\, 0 \,\vert\, n''-1,\, k'_y)\,
      [\kappa^{\ast}_1 \kappa''^{\ast}_1 + \kappa^{\ast}_3\kappa''^{\ast}_3]
      \Bigr\} e'_z \; ,\nonumber\\
   {\cal D}^{-+}_{n''n'}(\hbox{\bf k}) & = & \vsp
      J(k_x \,\vert\, n''-1,\, k'_y \,\vert\, n',\, -k''_y)\,
      [\kappa'_4\kappa''_3 + \kappa'_2\kappa''_1]\; e_-\nonumber\\
   && +\; J(k_x \,\vert\, n'',\, k'_y \,\vert\, n'-1,\, -k''_y )\,
     \vsp [\kappa'_3\kappa''_4 + \kappa'_1\kappa''_2]\; e_+\nonumber\\
   && -\; \Bigl\{ J(k_x \,\vert\, n'',\, k'_y \,\vert\, n',\, -k''_y )\,
      [\kappa'_4\kappa''_4 + \kappa'_2\kappa''_2] 
     - J(k_x \,\vert\, n''-1,\, k'_y \,\vert\, n'-1,\, -k''_y )\,
      [\kappa'_3\kappa''_3 + \kappa'_1\kappa''_1] \Bigr\} e_z \; ,\nonumber
\end{eqnarray}
where the polarization vector \teq{e_{\mu}=(0,e_x,e_y,e_z)} is specified
by \teq{e_{\pm}=e_x\pm ie_y} and \teq{e_z}, and similarly for the
final photon polarizations (primed).  The other three \teq{{\cal D}}s in
Eq.~(\ref{eq:mbweq25}) are not displayed here for brevity; they can be
obtained from those in Eq.~(\ref{eq:calD}) simply by the interchange
\teq{e_+\leftrightarrow e_-} of polarization components (and similarly
for primed components) and a relabelling of the \teq{J}s that produces
a correspondence \teq{{\cal D}^{q'q}_{n'n}(\hbox{\bf k}) \to {\cal
D}^{-q\, -q'}_{n\, n'}(\hbox{\bf k})}.

Several notations need to be identified.  First, the particles have
energies \teq{\erg}, and momentum components \teq{p_z} along the
field.  The energies \teq{\erg} and \teq{\erg_0} that appear here are,
respectively, with and without the parallel momentum \teq{p_z}:
\begin{equation}
   \erg\; =\;\sqrt{1+p_z^2+2nB}\quad ,\quad
   \erg_0\; =\;\sqrt{1+2nB \vphantom{p_z^2} }\quad ,
 \label{eq:erg}
\end{equation}
with \teq{n} denoting the Landau level quantum numbers, as usual.  The
other quantum number pertaining to the eigenstates of \teq{\mu_z} is
\teq{\sigma =\pm 1}, which signifies the spin state of the fermions
(\cite{mbw94} used the label \teq{\tau}; here the notation of Melrose
and Parle \cite{mpI} is preferred), and satisfies \teq{\mu_z\psi =
\sigma\erg_0\psi}.  It does not appear explicitly in \teq{\erg}, but is
embedded in the spinor coefficients \teq{\kappa_i}:
\begin{equation}
   \hbox{$ \pmatrix{\kappa_1 \vsp\cr  \kappa_2 \vsp\cr
         \kappa_3 \vsp\cr  \kappa_4 \vsp} $}\;\equiv\; 
   \hbox{$ \pmatrix{\delta_- & \delta_+ & 0 & 0 \vsp\cr
        \delta_+ & \delta_- & 0 & 0 \vsp\cr
        0 & 0 & \delta_- & -\delta_+ \vsp\cr
        0 & 0 & -\delta_+ & \delta_- \vsp} $}
   \hbox{$ \pmatrix{\sqrt{(\erg_0 +1)\, (\erg +\erg_0)} \vspm\cr
   -ip_z\,\sqrt{\dover{\erg_0-1}{\erg +\erg_0}} \vspl\cr
        p_z\,\sqrt{\dover{\erg_0+1}{\erg +\erg_0}} \vspl\cr
        i\,\sqrt{(\erg_0 -1)\, (\erg +\erg_0)} \vspm } $}\;\; ,
 \label{eq:kappadef}
\end{equation}
for \teq{\delta_+ = \delta_{\sigma, 1}} and \teq{\delta_- =
(1-\delta_{n,0})\,\delta_{\sigma, -1}} where the spin quantum number
\teq{\sigma} takes on two values except for the \teq{n=0} ground state
(zeroth Landau level), where only \teq{\sigma =1} is permissible.  Here
\teq{\delta_{i,j}} is the familiar Kronecker delta.  The primed coefficients
\teq{\kappa'_i} and \teq{\kappa''_i} are similarly defined in terms of
primed momenta and Landau level quantum numbers, subject to the momentum
conservation implicit in Eq.~(\ref{eq:mbweq25}).

The \teq{J} functions that appear in Eq.~(\ref{eq:calD}) are
integrals over the oscillator functions (Hermite polynomial products),
a form undeveloped in \cite{mbw94}.  Here, Eq.~(7.377) of \cite{gr80}
is employed to express these integrals analytically in terms of
generalized Laguerre polynomials, \teq{L^{n'-n}_{n}(x)} (see also Eq.~(47)
of \cite{mpI}):
\begin{equation}
   J(\alpha\vert n',\; \beta'\vert n,\;\beta )\; =\;
   \exp\Bigl(\, -i\,\dover{\alpha}{2B}\, [\beta +\beta']\Bigr)\,
   \Bigl\{ ie^{-i\psi}\Bigr\}^{n'-n}\, I_{n',n}\Bigl(\dover{\rho^2}{2B}\Bigr)
 \label{eq:Jeval}
\end{equation}
where \teq{\rho} (\teq{>0}) and the phase \teq{\psi} are introduced
for convenience of notation: \teq{\alpha =\rho\cos\psi} and
\teq{\beta'-\beta =\rho\sin\psi}.  Note that the \teq{\beta}s are
always \teq{k_y}s and \teq{\alpha} is always a \teq{k_x}.  Here the
\teq{I_{n',n}} functions follow the Sokolov and Ternov convention \cite{ST68}
up to a factor of \teq{n!}, being related to the \teq{{\bf J}}
functions of Melrose and Parle \cite{mpI}, and both are defined in
terms of the generalized Laguerre polynomials (see \cite{gr80}):
\begin{equation}
   I_{n',n}(x)\, =\, (-1)^{n'-n}\, I_{n,n'}(x)\, =\,
   {\bf J}_{n'-n}^{n}(x)\;\equiv\;\sqrt{\dover{n!}{n'!}}\;
   e^{-x/2}\, x^{(n'-n)/2}\, L^{n'-n}_{n}(x)\; ,\quad n' >n\; .
 \label{eq:IJdef}
\end{equation}
Values for \teq{n>n'} are obtained by interchanging indices, as
indicated.  Hereafter, the (modified) Sokolov and Ternov convention for
writing the Laguerre polynomials will be adopted.  Complex conjugation
of Eq.~(\ref{eq:Jeval}) can be used to establish the identity
\begin{equation}
   {\cal D}^{q'q}_{n'n}(\hbox{\bf k})\; =\; (-1)^{n'-n}\Bigl\{
   {\cal D}^{-q\, -q'}_{n\, n'}(\hbox{\bf k}) \Bigr\}^{\ast}\;\; ,
 \label{eq:calDsymm}
\end{equation}
noting that the \teq{\kappa} products are either purely imaginary or
real, for all choices of spin quantum numbers.  The three factors like
\teq{(-1)^{n'-n}} appearing in the second product of three \teq{{\cal
D}}s in Eq.~(\ref{eq:mbweq25}) cancel, leading to this product being
just the complex conjugate of the first three \teq{{\cal D}}s.  This
useful symmetry property clearly underlines the convenience of the
Sokolov and Ternov choice of wavefunctions when adopting real
components for the photon polarization.

The form of the contribution to the S-matrix element in
Eq.~(\ref{eq:mbweq25}) is {\it identical} to that for
\teq{S_{fi,1}^{(3)}} given in Eq.~(25) of Mentzel, Berg and Wunner
\cite{mbw94}.  In the same fashion, it can be found that
the expression derived here for \teq{S_{fi,2}^{(3)}} is absolutely
identical to Eq.~(26) of \cite{mbw94}, thereby providing confirmation
of their analytic developments; it can be obtained by using the
substitutions \teq{\hbox{\bf k}'\leftrightarrow \hbox{\bf k}''},
\teq{n'\leftrightarrow n''} and \teq{e'\leftrightarrow e''} in
Eq.~(\ref{eq:calD}).  All other \teq{S_{fi,j}^{(3)}}
contributions result from application of the cyclic permutations 
\begin{eqnarray}
   P_{+1}: &&
   k_{\mu}'',\; e_{\mu}''\;\to\; k_{\mu}',\; e_{\mu}'\quad ,\quad
   k_{\mu}',\; e_{\mu}'\;\to\; -k_{\mu},\; e_{\mu}^{\ast}\quad ,\quad
   -k_{\mu},\; e_{\mu}^{\ast}\;\to\; k_{\mu}'',\; e_{\mu}''\quad ;
 \nonumber\\[-5.5pt]
 \label{eq:perm}\\[-5.5pt]
   P_{-1}: &&
   -k_{\mu},\; e_{\mu}^{\ast}\;\to\; k_{\mu}',\; e_{\mu}'\quad ,\quad
   k_{\mu}',\; e_{\mu}'\;\to\; k_{\mu}'',\; e_{\mu}''\quad ,\quad
   k_{\mu}'',\; e_{\mu}''\;\to\; -k_{\mu},\; e_{\mu}^{\ast}\quad ,
 \nonumber
\end{eqnarray}
where \teq{k_{\mu}=(\omega, \hbox{\bf k})},
\teq{e_{\mu}=(0,e_x,e_y,e_z)}, etc.  Observe also that a minus sign and
the complex conjugation of the polarizations are always associated with
the initial photon since it is absorbed in the process.  Given these
permutations, the crossing symmetry for splitting is manifested in the
following relationship between the various terms like those in
Eq.~(\ref{eq:mbweq25}) that contribute to Eq.~(\ref{eq:totrate}):
\begin{equation}
   S^{(3)}_{fi,3}\; =\; P_{-1}\, S^{(3)}_{fi,2}\;\; ,\quad
   S^{(3)}_{fi,4}\; =\; P_{+1}\, S^{(3)}_{fi,1}\;\; ,\quad
   S^{(3)}_{fi,5}\; =\; P_{+1}\, S^{(3)}_{fi,2}\;\; ,\quad
   S^{(3)}_{fi,6}\; =\; P_{-1}\, S^{(3)}_{fi,1}\;\; ,
 \label{eq:crossymm}
\end{equation}
where the permutations act as operators.  This symmetry can be
expressed in a multitude of ways using the identities \teq{P_{+1}P_{-1}
= I = P_{-1}P_{+1}} and \teq{P_{\pm 1}^3 = I}. 

It is important to remark that the derivation of analytic forms by
Mentzel, Berg and Wunner is not the first in the literature relating to
S-matrix applications to photon splitting.  The papers by Melrose and
Parle \cite{mpI,mpII,mpIII} dealing with various aspects of QED in
strong magnetic fields, specifically from a wave dispersion/response
tensor approach, constructed the S-matrix element for splitting in
Eqs.~(46) and~(47) of \cite{mpII}, which incorporated the quadratic
vacuum response tensor given in Eq.~(36) of \cite{mpIII}.  This tensor
is obviously of a standard S-matrix Landau representation appearance.
Eqs.~(\ref{eq:mbweq25}) and~(\ref{eq:calD}) can be generated
directly (and also \teq{S_{fi,2}^{(3)}}) from the Melrose and Parle
evaluation after a modicum of algebra.  Hence, Eqs.~(\ref{eq:mbweq25})
and~(\ref{eq:calD}) here, and Eqs.~(25) and~(26) can be used as
reliable starting points for further S-matrix developments.

\subsection{Analytic Reduction: Summation over Spin States}
\label{sec:reduce}

The form in Eqs.~(\ref{eq:mbweq25}) and~(\ref{eq:calD}) is quite
cumbersome.  It can be simplified considerably by (i) specializing to
specific but representative directions of photon propagation and (ii)
analytically performing the summations over spin states \teq{\sigma},
\teq{\sigma'} and \teq{\sigma''}.  Restricting the photon motion to the
\teq{x}-direction yields photon motion perpendicular to the field:
since splitting is collinear in the non-dispersive limit discussed
earlier in this paper, it follows that \teq{k_z=k_z'=k_z''=0}.  This
choice dramatically simplifies coefficients of the Laguerre polynomials
in Eq.~(\ref{eq:calD}).  Without significant loss of generality,
setting \teq{k_y=k_y'=k_y''=0} removes nearly all of the phase factors
in the definition of the \teq{J}s in Eq.~(\ref{eq:Jeval}), leaving just
\teq{i^{n'-n}}.  Three such factors emerge in the triple product of
\teq{{\cal D}}s, leading to a factor of \teq{(-1)^{n''-n'}}.

The CP symmetry possessed by the splitting process becomes most evident
at this point, since it is now simple to derive the CP selection
rules.  The specification of \teq{k_y=k_y'=k_y''=0} and
\teq{k_z=k_z'=k_z''=0} yields only one possible component
of polarization perpendicular to the field, \teq{\eperp\,\equiv\,\erg_y
= -i\erg_+ = i\erg_-} and one conceivable component of polarization
parallel to the field, \teq{\epar\equiv\erg_z} (and similarly for
primed quantities).  The polarization (electric field) vector of the
photons is, of course, normal to the photon momentum vector, which
automatically spawns the notation for the two possible polarization
states:  \teq{\perp\; :\quad \eperp = 1,\quad \epar = 0}
and\ \teq{\parallel\; :\quad \eperp = 0,\quad \epar = 1}.  From the
presence of subtractions in the numerators of the integrands of
Eq.~(\ref{eq:mbweq25}) together with the complex conjugation property
in Eq.~(\ref{eq:calDsymm}) and the proportionality of the \teq{{\cal
D}}s to factors like \teq{i^{n'-n}}, it follows that only terms with an
odd number of \teq{\eperp} factors contribute to \teq{S_{fi,1}^{(3)}},
i.e. terms proportional to \teq{\epar\epar'\eperp''},
\teq{\epar\eperp'\epar''}, \teq{\eperp\epar'\epar''} and
\teq{\eperp\eperp'\eperp''}.  All other terms cancel identically to
zero.  By virtue of the permutation symmetries in Eq.~(\ref{eq:perm}),
this is also true for all other \teq{S_{fi,j}^{(3)}}.  It is then
trivial to deduce the CP selection rules for photon splitting, namely
that the only permitted transitions are
\begin{equation}
   \perp\;\to\;\perp\perp\quad ,\quad \perp\;\to\;\parallel\parallel
   \quad ,\quad \parallel\;\to\;\perp\parallel\quad .
 \label{eq:selrules}
\end{equation}
The three other splitting transitions all have S-matrix elements that
are exactly zero for collinear photon momenta, and hence are
forbidden.  This technique for CP selection rule derivation was
implemented in \cite{mbw94}.  These restrictions are simply
consequences of the charge conjugation (C) and parity (P) symmetries of
the splitting process, i.e.  relating to the transformations \teq{{\bf
k}\to -{\bf k}} and \teq{{\bf B}\to -{\bf B}}.  

The summation over the spin states \teq{\sigma}, \teq{\sigma'} and
\teq{\sigma''} (\teq{=\pm 1}) produces a dramatic simplification in the
appearance of the S-matrix elements.  Such spin summations act only on
the products of the \teq{\kappa_i}s that appear in Eq.~(\ref{eq:calD});
the algebra is lengthy but straightforward, being facilitated by
pairing \teq{S_{fi,j}^{(3)}} terms with denominators that differ only
in the sign of their photon energies.  The total splitting rate in
Eq.~(\ref{eq:totrate}) can be written in the form
\begin{equation}
   R_{e\to e'e''}\; =\; \dover{\fsc^3}{2\pi^2}\,
   \dover{mc^2}{\hbar} \int \dover{d\omega'}{\omega^2}\,
   \biggl\vert {\cal M}_{e\to e'e''} \biggr\vert^2\quad ,
 \label{eq:totratefin}
\end{equation}
where \teq{\omega''=\omega-\omega'} is implicitly understood from the
conservation of four-momentum.  While these rates will be
expressed for photon propagation normal to the uniform magnetic field,
the results for general photon obliquities \teq{\theta} to {\bf B} can
be obtained via a simple Lorentz transformation:
\teq{\omega\to\omega\sin\theta}, \teq{\omega'\to\omega'\sin\theta},
\teq{\omega''\to\omega''\sin\theta}, together with an extra
multiplicative factor of \teq{\sin\theta} applied to the rate in
Eq.~(\ref{eq:totratefin}).

The momentum dependence in the integrands of Eq.~(\ref{eq:mbweq25}) can
be simplified by forming sums of the products of energy denominators.
Separating such sums into real and imaginary parts via the
representation \teq{\Sigma_{\lambda,\mu}=
\Sigma_{\lambda,\mu}^{\hbox{\sevenrm R}} +i
\Sigma_{\lambda,\mu}^{\hbox{\sevenrm I}}}, leads to the definition
\begin{eqnarray}
\Sigma_{\lambda,\mu}^{\hbox{\sevenrm R}} &=& 
     \biggl\{ \dover{1}{(\erg'' +\erg+\omega')\,     
       (\erg'+\erg''+\omega)}
     +\dover{1}{(\erg'' +\erg-\omega')\, 
       (\erg'+\erg''-\omega)} \biggr\}    \nonumber\\[3pt]
  &+& \lambda\,\biggl\{ \dover{1}{(\erg' +\erg''-\omega)\,    
       (\erg+\erg'-\omega'')}
     +\dover{1}{(\erg' +\erg''+\omega )\, 
       (\erg+\erg'+\omega'')} \biggr\}   \label{eq:Sigmadef}\\[3pt]
  &+& \mu\,\biggl\{ \dover{1}{(\erg +\erg'+\omega'')\, 
       (\erg''+\erg-\omega')}
     +\dover{1}{(\erg +\erg'-\omega'')\,
       (\erg''+\erg+\omega )} \biggr\}    \quad ,\nonumber
\end{eqnarray}
of the real part, where \teq{\lambda} and \teq{\mu} assume the values
\teq{\pm 1}.  The imaginary part is not explicitly stated since it will
not be of use in the subsequent developments.  The momentum
integrations over these \teq{\Sigma_{\lambda,\mu}^{\hbox{\sevenrm R}}}
then assume one of the forms
\begin{eqnarray}
  & {\cal I}_{n} \; =\; \displaystyle{\int_{-\infty}^{\infty}} 
    \dover{p_z^{2n}\, dp_z}{\erg\erg'\erg''}
      \; \Sigma_{1,1}^{\hbox{\sevenrm R}}\;\; , &
   \nonumber\\[-6.5pt]
 \label{eq:calIdef}\\[-6.5pt]
   \quad {\cal J} \; =\; \int_{-\infty}^{\infty}
     \dover{dp_z}{\erg} \; \Sigma_{1,-1}^{\hbox{\sevenrm R}}\;\; , 
   & {\cal J}' \; =\; \displaystyle{\int_{-\infty}^{\infty}}
     \dover{dp_z}{\erg'} \; \Sigma_{-1,1}^{\hbox{\sevenrm R}}\;\; , 
         \vphantom{\Biggl(}
   & {\cal J}'' \; =\; \int_{-\infty}^{\infty}
     \dover{dp_z}{\erg''} \; \Sigma_{-1,-1}^{\hbox{\sevenrm R}}\nonumber
\end{eqnarray}
for \teq{n=0} or \teq{1}; generalizations to complex
\teq{\Sigma_{\lambda,\mu}} (relevant to calculating splitting rates
above pair threshold and near pair resonances) are routine.  These
manipulations yield the following compact forms for the \teq{{\cal
M}_{e\to e'e''}} coefficients in Eq.~(\ref{eq:totratefin}):
\begin{eqnarray}
 {\cal M}_{\pll} 
  & = &  -\dover{B}{4}  \sum_{n,n',n''}(-1)^{n''-n'} \biggl\{
   \vsp\sqrt{8n\, n'n''B^3}\; {\cal I}_0 \; 
          \triangle_1^{\pll}
  + \sqrt{2n''B}\; \Bigl[{\cal J}'' + {\cal I}_1 
            - {\cal I}_0 \Bigr] \; 
          \triangle_2^{\pll}\nonumber\\
  && \qquad\;\; +\sqrt{2n'B}\; \Bigl[{\cal J}' - {\cal I}_1 
            + {\cal I}_0 \Bigr] \;
          \triangle_3^{\pll}
      +\sqrt{2nB}\; \Bigl[{\cal J} + {\cal I}_1 
            + {\cal I}_0 \Bigr] \; 
          \triangle_4^{\pll}\vsp\biggl\}\nonumber\\
 {\cal M}_{\ppp} 
  & = &  -\dover{B}{4}  \sum_{n,n',n''}(-1)^{n''-n'} \biggl\{
   \vsp\sqrt{8n\, n'n''B^3}\; {\cal I}_0  \; \triangle_1^{\ppp}
  + \sqrt{2n''B}\; \Bigl[{\cal J}'' - {\cal I}_1 
            - {\cal I}_0 \Bigr] \; 
          \triangle_2^{\ppp}\nonumber\\[-6.5pt]
&& \label{eq:Mform}\\[-6.5pt]
  && \qquad\;\; +\sqrt{2n'B}\; \Bigl[{\cal J}' + {\cal I}_1 
            + {\cal I}_0 \Bigr] \;
          \triangle_3^{\ppp}
      +\sqrt{2nB}\; \Bigl[{\cal J} + {\cal I}_1 
            + {\cal I}_0 \Bigr] \; 
          \triangle_4^{\ppp}\vsp\biggl\}\nonumber\\
 {\cal M}_{\lpl} 
  & = &  -\dover{B}{4}  \sum_{n,n',n''}(-1)^{n''-n'} \biggl\{
   \vsp\sqrt{8n\, n'n''B^3}\; {\cal I}_0  \; 
          \triangle_1^{\lpl}
  + \sqrt{2n''B}\; \Bigl[{\cal J}'' + {\cal I}_1 
            - {\cal I}_0 \Bigr] \; 
          \triangle_2^{\lpl}\nonumber\\
  && \qquad\;\; +\sqrt{2n'B}\; \Bigl[{\cal J}' + {\cal I}_1 
            + {\cal I}_0 \Bigr] \;
          \triangle_3^{\lpl}
      +\sqrt{2nB}\; \Bigl[{\cal J} - {\cal I}_1 
            + {\cal I}_0 \Bigr] \; 
          \triangle_4^{\lpl}\vsp\biggl\}\;\; ,\nonumber
\end{eqnarray}
results that are to be used in conjunction with
Eq.~(\ref{eq:totratefin}).  The factor of \teq{-B/4} is introduced to
render the scaled amplitudes positive, and also to afford a direct
mapping onto limiting forms obtained \cite{bms96,bh97apj} by the
proper-time technique, as will become evident in
Section~\ref{sec:limits}.  The \teq{\Delta_i^{e\to e'e''}} are
differences of triple products of generalized Laguerre polynomials
(defined in Eq.~[\ref{eq:IJdef}]); for \teq{\pll}
\begin{eqnarray}
 \triangle_1^{\pll}\vsp
   = && I''_{n-1,n'-1}\, I'_{n'',n}\, I_{n',n''-1}
   - I''_{n,n'}\, I'_{n''-1,n-1}\, I_{n'-1,n''}\vsp\nonumber\\
 \triangle_2^{\pll}\vsp
   = && I''_{n-1,n'-1}\, I'_{n''-1,n-1}\, I_{n'-1,n''}
   - I''_{n,n'}\, I'_{n'',n}\, I_{n',n''-1}\vsp\nonumber\\[-5.5pt] 
 \label{eq:Ieq1}\\[-5.5pt]
 \triangle_3^{\pll}\vsp
   = && I''_{n-1,n'-1}\, I'_{n''-1,n-1}\, I_{n',n''-1}
   - I''_{n,n'}\, I'_{n'',n}\, I_{n'-1,n''}\vsp\nonumber\\
 \triangle_4^{\pll}\vsp
   = && I''_{n-1,n'-1}\, I'_{n'',n}\, I_{n'-1,n''}
   - I''_{n,n'}\, I'_{n''-1,n-1}\, I_{n',n''-1}\vsp\quad ,\nonumber
\end{eqnarray}
where the Sokolov and Ternov representation of the associated
Laguerre functions in Eq.~(\ref{eq:IJdef}) is used together with
the priming notation
\begin{equation}
   I_{n',n}\;\equiv\; I_{n',n}\biggl( \dover{\omega^2}{2B}\biggr)
        \quad ,\quad
   I_{n',n}'\;\equiv\; I_{n',n}\biggl( \dover{[\omega']^2}{2B}\biggr)
        \quad ,\quad
   I_{n',n}''\;\equiv\; I_{n',n}\biggl( \dover{[\omega'']^2}{2B}\biggr)
        \quad ,
 \label{eq:Iprimes}
\end{equation}
thereby aiding brevity.  For the \teq{\ppp} mode,
\begin{eqnarray}
 \triangle_1^{\ppp}\vsp
   = && I''_{n,n'-1}\, I'_{n'',n-1}\, I_{n',n''-1}
   - I''_{n-1,n'}\, I'_{n''-1,n}\, I_{n'-1,n''}\vsp\nonumber\\
 \triangle_2^{\ppp}\vsp
   = && I''_{n,n'-1}\, I'_{n''-1,n}\, I_{n'-1,n''}
   - I''_{n-1,n'}\, I'_{n'',n-1}\, I_{n',n''-1}\vsp\nonumber\\[-5.5pt]
 \label{eq:Ieq2}\\[-5.5pt]
 \triangle_3^{\ppp}\vsp
   = && I''_{n,n'-1}\, I'_{n''-1,n}\, I_{n',n''-1}
   - I''_{n-1,n'}\, I'_{n'',n-1}\, I_{n'-1,n''}\vsp\nonumber\\
 \triangle_4^{\ppp}\vsp
   = && I''_{n,n'-1}\, I'_{n'',n-1}\, I_{n'-1,n''}
   - I''_{n-1,n'}\, I'_{n''-1,n}\, I_{n',n''-1}\vsp\quad ,\nonumber
\end{eqnarray}
and the results for the \teq{\lpl} mode are not explicitly stated since
they can be obtained by exploiting crossing symmetries:  the inverse of
the permutation in Eq.~(\ref{eq:permsymm}) yields the transformation
\teq{\triangle_1^{\pll}\to -\triangle_1^{\lpl}},
\teq{\triangle_2^{\pll}\to \triangle_3^{\lpl}},
\teq{\triangle_3^{\pll}\to -\triangle_4^{\lpl}},
\teq{\triangle_4^{\pll}\to \triangle_2^{\lpl}}.  The
\teq{\triangle_i^{e\to e'e''}} can alternatively be expressed using
the \teq{{\bf J}^a_b} functions of Melrose and Parle as in
\cite{wbm98}.  Note that the potential subtlety of having to include
factors of \teq{1/2} for some contributions from ground intermediate
states is eliminated by the specific choice of the Sokolov and Ternov
wavefunctions.

The comparative simplicity of the reduced form of the S-matrix element
relative to Eq.~(\ref{eq:mbweq25}) is both notable and comforting.
Unlike Eqs.~(25) and~(26) of \cite{mbw94}, this developed form of the
splitting S-matrix element has an appearance familiar to S-matrix
applications of QED in the Landau representation to strongly-magnetized
systems, with products of generalized Laguerre polynomials multiplied
by simple combinations of energies and momentum components.  Examples
of previous work bearing such familiar forms focus largely on
lower-order QED processes and include studies of synchrotron radiation
\cite{ST68,hrw82}, single photon pair creation \cite{dh83,ST68}, and
vacuum \cite{ms77} and plasma \cite{psy80} polarization.

For the purposes of the analysis in the next section, it is pertinent
to define the cyclic permutations
\begin{eqnarray}
   && \omega   \;\to\; -\omega''\;\; ,\quad
      \omega'  \;\to\; -\omega\;\; ,\quad
      \omega'' \;\to\;  \omega'\;\; ,\nonumber\\[-6.5pt]
 \label{eq:permsymm}\\[-6.5pt]
   && n   \;\to\; n''\;\; ,\quad
      n'  \;\to\; n\;\; ,\quad
      n'' \;\to\; n'\;\; ,\nonumber
\end{eqnarray}
in the spirit of the \teq{P_{+1}} permutation in Eq.~(\ref{eq:perm}).
These permutations will appear repeatedly in the developments below,
and lead to the following transformation properties of
Eq.~(\ref{eq:Sigmadef}):
\begin{equation}
 \Sigma_{-1,-1}^{\hbox{\sevenrm R}} \;\to\;
      -\Sigma_{-1,1}^{\hbox{\sevenrm R}}\;\; ,\quad 
 \Sigma_{-1,1}^{\hbox{\sevenrm R}} \;\to\;
       \Sigma_{1,-1}^{\hbox{\sevenrm R}}\;\; ,\quad 
 \Sigma_{1,-1}^{\hbox{\sevenrm R}} \;\to\;
      -\Sigma_{-1,-1}^{\hbox{\sevenrm R}}\;\; ,\quad 
 \label{eq:Sigmasymm}
\end{equation}
with \teq{\Sigma_{1,1}^{\hbox{\sevenrm R}}} being invariant, symmetries
that are consequences of the arrangements of electron and positron
propagators in the Feynman diagram for splitting.  These translate into
obvious mappings between \teq{{\cal J}}, \teq{{\cal J}'} and \teq{{\cal
J}''} and an invariance of the \teq{I_n}.  It is also easily seen that
under this cyclic permutation, the factor in braces in the summation
for \teq{{\cal M}_{\ppp}} is invariant, while the equivalent factor in
the summation for \teq{{\cal M}_{\lpl}} maps over (up to a minus sign)
to the factor in braces in the \teq{{\cal M}_{\pll}} summation.  As
will become evident in Section~\ref{sec:limits}, the remaining powers
of \teq{-1} in the summations do not provide any unsatisfactory
interference in the limits of low photon energy (\teq{\omega\ll 1}) and
high fields (\teq{B\gg 1}), so that permutation symmetry can be
extended to the total amplitudes in these specific parameter regimes.

\subsection{Analytic Reduction: Integration over Parallel Momentum}
\label{sec:pzint}

Further analytic development is not only possible, but also desirable,
given that the integrations over the momentum \teq{p_z} parallel to the
field can be expressed compactly in terms of elementary functions.
Such tractability facilitates both numerical evaluations and the
derivation of asymptotic limits.   In proceeding, since results are
sought at energies sufficiently remote from pair creation resonances,
the imaginary parts of the denominators in the \teq{\Sigma_{\lambda,
\mu}} are dropped in all further considerations, i.e., we consider only
the functions \teq{\Sigma_{\lambda,\mu}^{\hbox{\sevenrm R}}=\hbox{Re}
\Sigma_{\lambda,\mu}}.

It turns out that carefully-constructed contour integrations in the
complex \teq{p_z} plane do not facilitate the \teq{p_z} integrations.
Hence the first step in integrating over \teq{p_z} is effected by the
more cumbersome and less elegant approach of completing the squares and
rationalizing the denominators using products of factors like
\teq{(\erg'\pm\erg''\pm\omega)}.  These factors define poles
\teq{p_{ij}} of the \teq{p_z} integration for \teq{i} and \teq{j} being
some combination of \teq{n}, \teq{n'} and \teq{n''}.  Such poles fall
into two types: pair creation ones (e.g. see \cite{dh83}) that
contribute only above pair threshold, due to the structure of the
splitting rate, and cyclotronic ones that must be considered below pair
threshold.  The appearance of such cyclotronic poles is an artifact of the
rationalization of denominators, so that they are really pseudo-poles
of the subsequent analysis; a consistency check on the algebra is that
the S-matrix element be effectively continuous across them.  It is
convenient to define energies that correspond to the \teq{p_{ij}}
poles:
\begin{equation}
   \erg_{nn'}\; =\; \dover{(\omega'')^2 + {\cal N}-{\cal N}'}{2\omega''}
   \quad , \quad
   \erg_{n'n''}\; =\; -\dover{\omega^2 + {\cal N}'-{\cal N}''}{2\omega}
   \quad , \quad
   \erg_{n''n}\; =\; \dover{(\omega')^2 + {\cal N}''-{\cal N}}{2\omega'}
 \label{eq:ergij}
\end{equation}
and three others paired with these, which are obtained via the
relations \teq{\erg_{n'n} + \erg_{nn'} = \omega''}, \teq{\erg_{n''n'} +
\erg_{n'n''} = -\omega} and \teq{\erg_{nn''} + \erg_{n''n} = \omega'}.
Here the notation
\begin{equation}
   {\cal N}\; =\; 1+2nB\;\; ,\quad  {\cal N}'\; =\; 1+2n'B\;\; ,
   \quad  {\cal N}''\; =\; 1+2n''B
 \label{eq:Ndef}
\end{equation}
is used for the purposes of abbreviation.  Observe that, taking
advantage of the subjectivity of such definitions, a minus sign appears
in front of the expression for \teq{\erg_{n'n''}}, a choice that
preserves symmetries induced by the mapping in Eq.~(\ref{eq:permsymm})
in the results that follow.  These definitions spawn the following
useful identities for the momentum poles:
\begin{eqnarray}
   p_{nn'}^2 &=&  \erg_{nn'}^2 - {\cal N}
       \; =\; \erg_{n'n}^2  - {\cal N}' \nonumber\\
   p_{n'n''}^2 &=& \erg_{n'n''}^2 - {\cal N}'
       \; =\; \erg_{n''n'}^2  - {\cal N}'' 
 \label{eq:poleident} \\
   p_{n''n}^2 &=& \erg_{n''n}^2 - {\cal N}''
       \; =\; \erg_{nn''}^2  - {\cal N} \;\; ,
   \nonumber
\end{eqnarray}
which immediately imply the possibility of poles along the imaginary
axis.  In fact, \teq{p_{nn'}^2\geq -\hbox{min}\{ {\cal N},\, {\cal
N}'\} }, with equality for \teq{\omega''=\vert {\cal N}-{\cal
N}'\vert^{1/2}}, and likewise for the other poles.   Note that for the
one-vertex calculations of cyclotron emission and single photon pair
creation and annihilation, the requirement that such poles be real,
corresponding to real components of particle momenta on external lines,
is precisely what generates thresholds (e.g. \cite{dh83}) and kinematic
cutoffs (e.g. \cite{hrw82}) for transitions involving various states.

The rationalization of the denominators yields relatively compact 
decompositions for these sums, after much cancellation and simplification.
They take the form 
\begin{equation}
   \Sigma_{\lambda,\mu}^{\hbox{\sevenrm R}}\; =\; c_{\lambda,\mu} +
   \dover{2}{W}\; \Bigl\{ t_{\lambda,\mu}^{\erg\erg'}\, \erg\erg' +
   t_{\lambda,\mu}^{\erg'\erg''}\, \erg'\erg'' +
   t_{\lambda,\mu}^{\erg''\erg}\, \erg''\erg \Bigr\}\;\; ,
 \label{eq:Sigmacoeffdef}
\end{equation}
the simplicity of which is contingent upon the energy-conservation
restriction \teq{\omega''=\omega -\omega'}.  Here
\begin{equation}
   W \; =\; \omega\omega'\omega''
        +\omega{\cal N}-\omega'{\cal N}'-\omega''{\cal N}'\;\; .
 \label{eq:Wdef}
\end{equation}
Identities such as \teq{W=-2\omega\omega' (\erg_{n''n'}+\erg_{n''n})}
prove useful in the ensuing analysis.  The \teq{c_{\lambda,\mu}} and
\teq{t_{\lambda,\mu}} coefficients assume simple forms when expressed
as partial fractions.  Consider first the result for
\teq{\Sigma_{1,1}^{\hbox{\sevenrm R}}}, which has the coefficients
\begin{eqnarray}
   c_{1,1} &=& 0\nonumber\\
   t_{1,1}^{\erg\erg'} &=&
     \dover{\erg_{n''n}}{p_{n''n}^2 - p_z^2} +  
     \dover{\erg_{n''n'}}{p_{n'n''}^2 - p_z^2} \nonumber\\[-5.5pt]
 \label{eq:Sigma11coeff}\\[-5.5pt]
   t_{1,1}^{\erg'\erg''} &=& 
     \dover{\erg_{nn''}}{p_{n''n}^2 - p_z^2} +  
     \dover{\erg_{nn'}}{p_{nn'}^2 - p_z^2} \nonumber\\[2pt]
   t_{1,1}^{\erg''\erg} &=& 
     \dover{\erg_{n'n}}{p_{nn'}^2 - p_z^2} +  
     \dover{\erg_{n'n''}}{p_{n'n''}^2 - p_z^2} \;\; . \nonumber
\end{eqnarray}
Observe that a cyclic symmetry is immediately apparent:
\teq{\Sigma_{1,1}^{\hbox{\sevenrm R}}} is invariant under the
permutation in Eq.~(\ref{eq:permsymm}), as is evident from its original
definition in Eq.~(\ref{eq:Sigmadef}).  Similarly, the algebraic
developments yield coefficients for the
\teq{\Sigma_{-1,-1}^{\hbox{\sevenrm R}}} sum, which appears in the
\teq{\erg\erg'\Delta_2} terms, as
\begin{eqnarray}
   c_{-1,-1} &=& \dover{1}{\omega\omega'} - \dover{2}{W}\,\biggl\{ 
     \dover{\erg_{n''n'}\erg_{n'n''}}{p_{n'n''}^2 - p_z^2} 
   \;[\erg_{n''n'}+\omega']
   + \dover{\erg_{n''n}\erg_{nn''}}{ p_{n''n}^2 - p_z^2}
    \;[\erg_{n''n}-\omega ]\,\biggr\} \nonumber\\
   t_{-1,-1}^{\erg\erg'} &=& 0 \nonumber\\[-5.5pt] 
  && \label{eq:Sigmam1m1coeff}\\[-5.5pt]
   t_{-1,-1}^{\erg'\erg''} &=& 
     - \dover{\erg_{n''n'}+\omega'}{p_{n'n''}^2 - p_z^2} -  
     \dover{\erg_{nn'}}{p_{nn'}^2 - p_z^2} \nonumber\\[2pt]
   t_{-1,-1}^{\erg''\erg} &=& 
     - \dover{\erg_{n''n}-\omega}{p_{n''n}^2 - p_z^2} -  
     \dover{\erg_{n'n}}{p_{nn'}^2 - p_z^2} \;\; . \nonumber
\end{eqnarray}
The coefficients for the sum \teq{\Sigma_{-1,1}^{\hbox{\sevenrm R}}}
that appears in the \teq{\erg\erg''\Delta_3} terms and the coefficients
for the sum \teq{\Sigma_{1,-1}^{\hbox{\sevenrm R}}} that appears in the
\teq{\erg'\erg''\Delta_4} terms are similar: there is little need to
state them explicitly, since the coefficients possess a
relationship to each other due to the permutation symmetry enunciated
in Eq.~(\ref{eq:Sigmasymm}).   

Given these decompositions, it is now fairly straightforward to
evaluate the integrations over \teq{p_z}, expressing them in terms of
the an elementary function \teq{f} with real arguments \teq{\erg_{ij}}:
\begin{equation}
   f({\cal N},\, {\cal E} )\; \equiv\;
      P\!\! \int^{\infty}_{-\infty} \dover{dp_z}{\sqrt{{\cal N} +p_z^2}}
      \, \dover{{\cal E}}{{\cal E}^2-{\cal N}-p_z^2} 
   \; =\; \cases{ \vphantom{\Biggl(}
       \dover{1}{\sqrt{{\cal E}^2-{\cal N}}}\;\log_e \Biggl\vert
         \dover{{\cal E} +\sqrt{{\cal E}^2-{\cal N}}}{
         {\cal E} -\sqrt{{\cal E}^2-{\cal N}}}\; \Biggr\vert\quad ,
       & if \teq{{\cal E}^2\, >\, {\cal N}\;\;},\cr
       -\dover{2}{\sqrt{{\cal N}-{\cal E}^2}}\; \arctan\biggl\{
          \dover{{\cal E}}{\sqrt{{\cal N}-{\cal E}^2}}\,
          \biggr\} \vphantom{\Biggl(}\;\; ,
       & if \teq{0 \, <\, {\cal E}^2 \, <\, {\cal N} } \cr}
 \label{eq:ffuncdef}
\end{equation}
for real \teq{{\cal E}}.  The identity \teq{\arctan
z=(1/2i)\log_e[(1+iz)/(1-iz)]} with \teq{z=-{\cal E}/\sqrt{{\cal N}-
{\cal E}^2}} has been used to map across the singularities at
\teq{{\cal E}=\pm\sqrt{{\cal N}}} (cyclotronic below pair threshold)
and guarantee bounded and continuous behaviour of \teq{f({\cal N},\,
{\cal E} )/ {\cal E}} at \teq{{\cal E}=0}.  The integral identity in
Eq.~(\ref{eq:ffuncdef}) can be established quickly with the aid of
result 3.513.2 in \cite{gr80}, using the substitution
\teq{p_z=\sqrt{{\cal N}}\sinh t} and partial fractions.  Note that real
values (either positive or negative) of \teq{{\cal E}} are guaranteed
by the formalism here, with \teq{{\cal E}=0} being improbable due to
the discreteness of the quantum numbers \teq{n}, \teq{n'} and
\teq{n''}.

The integration of the coefficients \teq{{\cal I}_0} of the
\teq{\Delta_1} terms for each of the polarization modes are then
straightforward, and the identities in Eq.~(\ref{eq:poleident}) can be
used to advantage.  Similar terms appear in the \teq{{\cal I}_1}
integrations of parts of the coefficients of the other\teq{\Delta_i}
terms, which also possess integrands with terms proportional to
\teq{1/\erg}, \teq{1/\erg'} and \teq{1/\erg''} that formally lead to
divergences that cancel each other (an artifice introduced by the
rationalization of the denominators).  Using partial fractions, the
divergent contributions can be written as integrals over the finite
range \teq{-p\leq p_z\leq p}, rearranging to subtract off
exactly-cancelling terms, and then taking the limit as
\teq{p\to\infty}.  Similar manipulations are used for the \teq{{\cal
J}''} integration over \teq{\Sigma_{-1,-1}^{\hbox{\sevenrm R}}}, where
again the leading order terms are individually divergent yet
collectively convergent.  Partial fractions can again be used to enable
rearrangements and separate the divergent terms, which are then
integrated over finite ranges as with the \teq{{\cal I}_1} evaluation.
The results are encapsulated in the identities
\begin{eqnarray}
  {\cal I}_0  & = & \dover{2}{W}\; \biggl\{
                 {\cal F}_{nn'}
               + {\cal F}_{n'n''}
               + {\cal F}_{n''n} \biggr\}\;\; , \nonumber\\
  {\cal I}_1  & = & {\cal L} + \dover{2}{W}\; \biggl\{
                 p_{nn'}^2\, {\cal F}_{nn'} 
               + p_{n'n''}^2\, {\cal F}_{n'n''}
               + p_{n''n}^2\, {\cal F}_{n''n} \biggr\}\;\; , 
 \label{eq:Sigmaint}\\
  {\cal J}'' & = &  {\cal L} - \dover{2}{W}\; \biggl\{ 
                 \erg_{nn'}\erg_{n'n}\, {\cal F}_{nn'}
               + \erg_{n'n''} (\erg_{n''n'}+\omega') \, {\cal F}_{n'n''}
               + \erg_{nn''} (\erg_{n''n}-\omega) \, {\cal F}_{n''n}
       \biggr\} \;\; ,\nonumber
\end{eqnarray}
where
\begin{equation}
  {\cal F}_{nn'} = f({\cal N},\, \erg_{nn'})
            + f({\cal N}',\, \erg_{n'n}) , \quad
  {\cal F}_{n'n''} = f({\cal N}',\, \erg_{n'n''}) 
            + f({\cal N}'',\, \erg_{n''n'}) , \quad
  {\cal F}_{n''n} = f({\cal N}'',\, \erg_{n''n})
            + f({\cal N},\, \erg_{nn''}) , 
 \label{eq:calFdef}
\end{equation}
and
\begin{equation}
  {\cal L}\; =\; \dover{1}{\omega'\omega''}\,\log_e{\cal N}
                    - \dover{1}{\omega''\omega}\,\log_e{\cal N}'
                    - \dover{1}{\omega\omega'}\,\log_e{\cal N}''\;\; .
 \label{eq:calLdef}
\end{equation}
No further integration is necessary: the cyclic permutations in
Eq.~(\ref{eq:permsymm}) can be used to quickly derive expressions for
\teq{{\cal J}'} and \teq{{\cal J}} from Eq.~(\ref{eq:Sigmaint}).

At this point, it is salient to remark that the divergences at
\teq{{\cal E}^2= {\cal N}} in the functions \teq{f({\cal N}, {\cal E})}
pose no problem for the integral evaluations in
Eqs.~(\ref{eq:Sigmaint}), because these functions always appear two at
a time.  Below the pair threshold, these divergences are cyclotronic in
nature, being encountered when \teq{\omega\to \vert \sqrt{{\cal N}'}
-\sqrt{{\cal N}''}\vert} or for similar circumstances for the other
photon energies.  As \teq{\omega} tends to such a limit, for example,
we observe that \teq{\erg_{n'n''}\to\sqrt{{\cal N}'}} and
\teq{\erg_{n''n'}\to -\sqrt{{\cal N}''}} when \teq{{\cal N}'>{\cal
N}''} (without loss of generality).  This opposition of signs
guarantees cancellation of divergences when the \teq{\arctan} form of
\teq{f({\cal N}, {\cal E})} is used (\teq{\arctan (1/z)\to \pi/2 - z}
as \teq{z\to 0}), so that continuity across cyclotron
``pseudo-resonances'' emerges naturally from Eq.~(\ref{eq:Sigmaint}),
consistent with the continuity of the \teq{\Sigma_{\lambda,
\mu}^{\hbox{\sevenrm R}}} functions.  Continuity across pair resonances
does not arise above pair threshold, so that true divergences emerge.

The incorporation of Eq.~(\ref{eq:Sigmaint}) into the scaled matrix
elements in Eq.~(\ref{eq:Mform}) constitutes the final product of the
general analytic developments in this paper, providing rates valid for
all energies below pair threshold (and applicable for non-resonant
energies above threshold), and for photon propagation normal to the
uniform magnetic field.  They are eminently suitable for numerical
computations, having improved upon the analytic formalism of Mentzel,
Berg and Wunner \cite{mbw94} (i.e. Eq.~[\ref{eq:mbweq25}]) by
performing the summations of the spin states and integration over the
momenta parallel to the field that are associated with the electron
propagators.  Such developments are prudent prior to numerical
evaluations due to the large degree of cancellation in these sums and
integrations.

\section{ASYMPTOTIC LIMITS FOR HIGH \teq{B} OR SMALL \teq{\omega}}
\label{sec:limits}

A fruitful extension of this analysis is the exploration of the
simplification of the scattering amplitudes and rates in two particular
asymptotic regimes, namely the limit of highly supercritical fields,
\teq{B\gg 1}, and the specialization to photon energies well below
threshold, i.e. \teq{\omega\ll 1}.  The benefits of such an
investigation are twofold.  First, it provides the first unequivocal
analytic demonstration of the equivalence of splitting results from the
S-matrix formulation in the Landau representation and effective
Lagrangian/proper-time results from Schwinger-type formalisms in
well-defined parameter regimes.  In doing so, it serves as a powerful
check on the developments here.  Second, in the \teq{\omega\ll 1} case,
it identifies a new, satisfyingly compact representation of the
scattering amplitudes in terms of special functions that leads to an
efficient means of computation.

These two parameter regimes are encompassed under the single limit
\teq{\omega^2\ll 1+2B}, which thereby identifies the appropriate series
expansion of the generalized Laguerre polynomials that appear in the
amplitudes.  For small arguments \teq{x}, the leading order terms in
the series for \teq{I_{n',n}(x)} can be found in the Appendix of
\cite{mpI}.  Given that \teq{n}, \teq{n'} and \teq{n''} cluster in a
manner such that \teq{\vert n'-n\vert\sim \vert n''-n \vert\sim 1},
this series converges rapidly provided \teq{nx\ll 1}.  Hence
\teq{n\omega^2/(2B)} actually represents the true expansion parameter
here, with \teq{\omega'} and \teq{\omega''} being similarly bounded.
The leading order terms of such expansions for the \teq{\Delta_i^{e\to
e'e''}} are linear in the photon energies, while the next higher order
terms are cubic; a more detailed exposition can be found in Weise,
Baring and Melrose \cite{wbm98}.  The series for the integrations of
\teq{p_z}, namely \teq{{\cal I}_0}, \teq{{\cal I}_1}, \teq{{\cal J}''},
\teq{{\cal J}'} and \teq{{\cal J}} (which do not depend on the
polarization mode) are expansions in \teq{\omega^2/(1+2B)} rather than
\teq{\omega^2/(2B)}.  They are independent of photon energy to leading
order, with a quadratic scaling with energy to next order.  The series
for \teq{\omega^2\ll 1+2B} possess logarithmic character in the quantum
numbers in situations when no two of them are equal (i.e. \teq{{\cal
N}\neq {\cal N}'\neq {\cal N}''\neq {\cal N}}):
\begin{eqnarray}
  {\cal I}_0  & \approx &
     \dover{4\log_e {\cal N}}{
            ({\cal N}-{\cal N}') ({\cal N}''-{\cal N})} +
     \dover{4\log_e {\cal N}'}{
            ({\cal N}'-{\cal N}'') ({\cal N}-{\cal N}')} +
     \dover{4\log_e {\cal N}''}{
            ({\cal N}''-{\cal N}) ({\cal N}'-{\cal N}'')}\;\; ,
   \nonumber\\[-5.5pt]
 \label{eq:Iapprox}\\[-5.5pt]
  {\cal I}_1 \;\approx\; - {\cal J}'' & \approx & 
     -\dover{2 {\cal N} \log_e {\cal N}}{
            ({\cal N}-{\cal N}') ({\cal N}''-{\cal N})} -
      \dover{2 {\cal N}' \log_e {\cal N}'}{
            ({\cal N}'-{\cal N}'') ({\cal N}-{\cal N}')} -
      \dover{2 {\cal N}'' \log_e {\cal N}''}{
            ({\cal N}''-{\cal N}) ({\cal N}'-{\cal N}'')}\;\; ,
   \nonumber
\end{eqnarray}
and additionally involve inverse trigonometric functions when 
two \teq{n}s (e.g. for \teq{{\cal N}={\cal N}'}) are in fact equal:
\begin{eqnarray}
  {\cal I}_0  & \approx & 
     \dover{4}{({\cal N}-{\cal N}'')^2}\, 
        \log_e \dover{{\cal N}}{{\cal N}''} -
     \dover{4}{{\cal N} ({\cal N}-{\cal N}'')} \;
        Q\Bigl( \dover{\omega''}{2\sqrt{{\cal N}}} \Bigr)\;\; ,
   \nonumber\\[-5.5pt]
 \label{eq:Iapprox2}\\[-5.5pt]
  {\cal I}_1  & \approx & -
     \dover{2 {\cal N}''}{({\cal N}-{\cal N}'')^2}\,
          \log_e \dover{{\cal N}}{{\cal N}''} -
     \dover{2}{{\cal N}-{\cal N}''} +
     \dover{4{\cal N}-(\omega'')^2}{{\cal N} ({\cal N}-{\cal N}'')} \;
        Q\Bigl( \dover{\omega''}{2\sqrt{{\cal N}}} \Bigr)
      \;\approx\; \dover{4}{({\cal N}-{\cal N}'')} \;
        Q\Bigl( \dover{\omega''}{2\sqrt{{\cal N}}} \Bigr) - {\cal J}''\;\; ,
   \nonumber
\end{eqnarray}
where
\begin{equation}
     Q(x)\; =\; \dover{\arcsin x}{x\sqrt{1-x^2}}
 \label{eq:qdef}
\end{equation}
and the identity \teq{\arcsin x = \arctan [x/\sqrt{1-x^2}]} has been
invoked.  This retention of the inverse trigonometric functions is
particularly relevant for determining the high \teq{B} limiting forms
of the scattering amplitudes.  Relations similar to
Eq.~(\ref{eq:Iapprox2}) exist for \teq{{\cal N}={\cal N}'} and
\teq{{\cal N}'={\cal N}''}, obtained by the cyclic permutations through
\teq{{\cal N}}s and photon energies.  The lengthier higher order
(quadratic) terms are not explicitly stated for the sake of brevity.
This concludes the preamble that guides the reader in the subsequent
specializations.

\subsection{The Special Case of \teq{B\gg 1}}
\label{sec:Bgg1}

\begin{figure}
%
%\vspace{0.4truecm}
\centerline{\psfig{figure=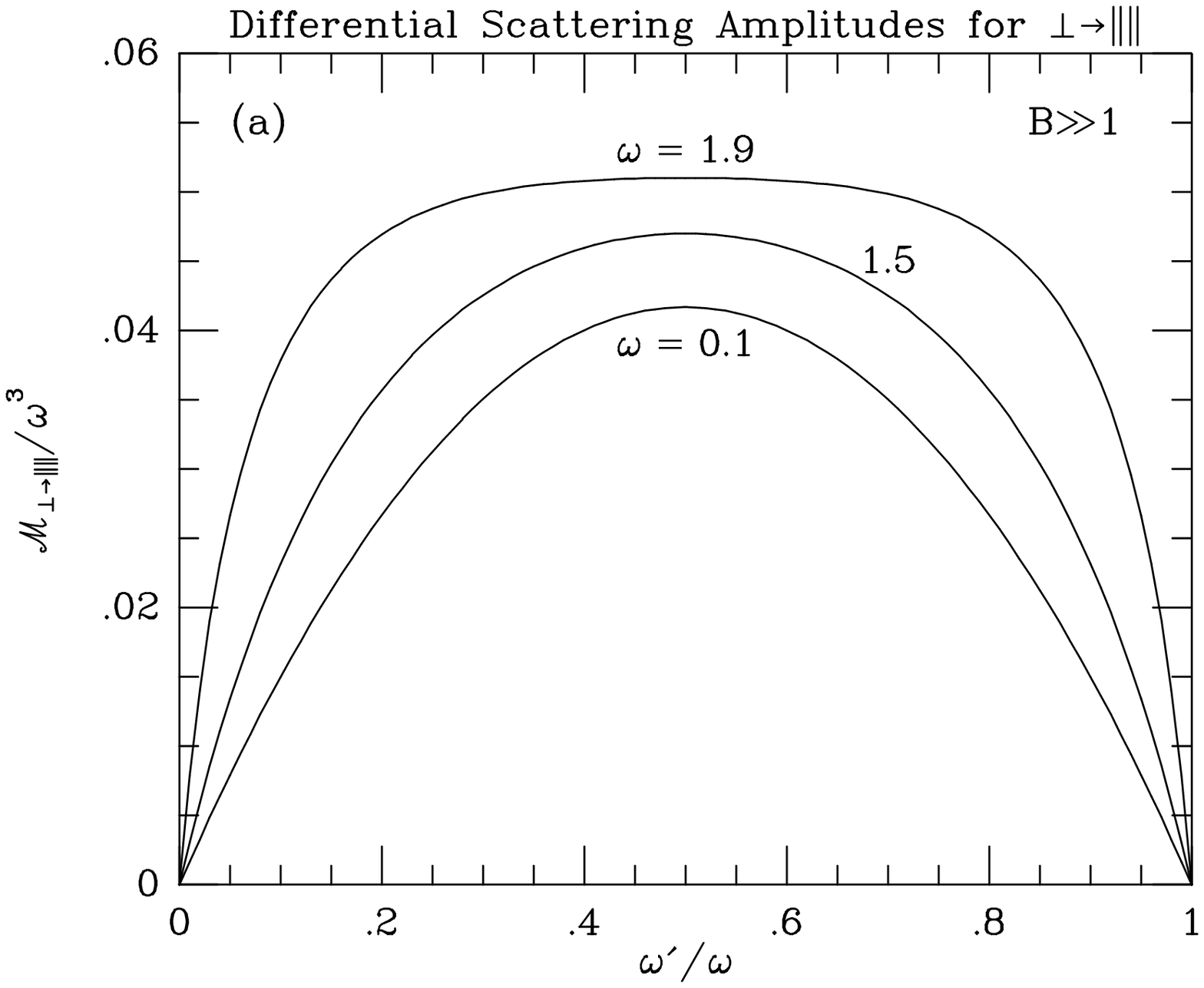,width=8.5cm}\hskip 0.5truecm
            \psfig{figure=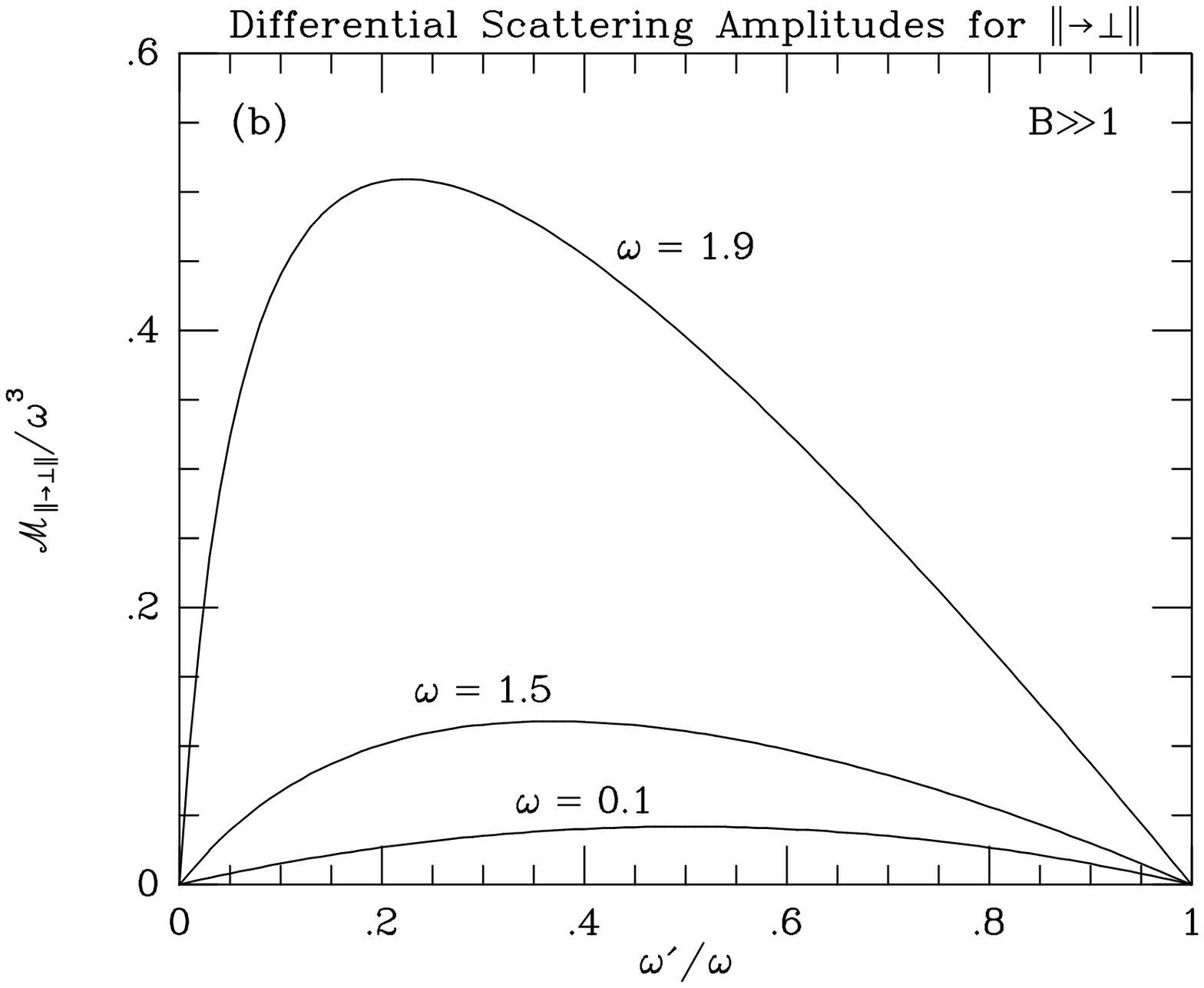,width=8.5cm}}
\caption{The dependence of the scattering amplitudes for \teq{B\gg 1},
scaled by \teq{\omega^3}, on the fractional energy \teq{\omega'/\omega}
of one of the produced photons, for three different incident photon
energies \teq{\omega} (in units of \teq{mc^2}), as labelled.  Only the
two polarization modes with amplitudes asymptotically independent of
\teq{B} (in units of \teq{B_{\rm c}}) in this ultra-quantum limit are
depicted, namely (a) \teq{\perp\to\parallel\,\parallel} and (b)
\teq{\parallel\to\perp\,\parallel}; their functional forms are given in
Eqs.~(\ref{eq:MperptoparparBgg1}) and~(\ref{eq:MpartoperpparBgg1}),
respectively.  The shape of the amplitude curves for \teq{\ppp} is
independent of \teq{\omega} and is very close to that of the
\teq{\omega=0.1} curves in panels (a) and (b).  While the
\teq{\perp\to\parallel\,\parallel} curves are necessarily symmetric
about \teq{\omega'=\omega/2}, asymmetry is present in the
\teq{\parallel\to\perp\,\parallel} case where \teq{\omega'} represents
the final photon of \teq{\perp} polarization.  Note that the magnitude
of \teq{{\cal M}_{\parallel\to\perp\,\parallel}} diverges as pair
threshold \teq{w=2} is approached.}
 \label{fig:scattampBgg1}
\end{figure}

This regime is of particular relevance to the study of magnetars such
as soft gamma repeaters.  For the two modes \teq{\pll} and \teq{\lpl},
only the leading order terms for the \teq{\Delta_i} and the momentum
integrals presented in Eqs.~(\ref{eq:Iapprox}) and~(\ref{eq:Iapprox2})
are required.  Consider first the reduction of \teq{{\cal M}_{\pll}}.
Here the \teq{\Delta_2} and \teq{\Delta_3} terms contribute leading
order terms only through \teq{n''=1}, \teq{n=n'=0} and \teq{n'=1},
\teq{n=n''=0} cases, respectively, where it is necessary to use the
full forms in Equation~(\ref{eq:Iapprox2}), and inverse trigonometric
functions appear through the \teq{Q(x)} function, which assumes the
arguments \teq{x=\omega'/2} and \teq{x=\omega''/2}.  A similar
\teq{n=1}, \teq{n'=n''=0} term is identically equal to zero by virtue
of the \teq{\Delta_4} factor.  The contributions from the
\teq{\Delta_1} and \teq{\Delta_4} terms possess an entirely different
character, being infinite summations over \teq{n}, with the values of
\teq{n'} and \teq{n''} being constrained by \teq{\vert n'-n \vert +
\vert n''-n \vert \leq 1}, producing five groupings of the indices.
The series is evaluated by truncating the sum at \teq{n\leq k},
relabelling one of the logarithmic terms, and then taking the limit
\teq{k\to\infty}.  The net result is (for \teq{\omega < 2})
\begin{equation}
   {\cal M}_{\pll}\; \approx \; 
   \dover{4\omega'}{\omega''\sqrt{4-(\omega'')^2}}\,\arcsin
      \Bigl(\dover{\omega''}{2}\Bigr) +
   \dover{4\omega''}{\omega'\sqrt{4-(\omega')^2}}\,\arcsin 
      \Bigl(\dover{\omega'}{2}\Bigr) - \omega
   \quad , \quad B\gg 1\;\; ,
 \label{eq:MperptoparparBgg1}
\end{equation}
which, when combined with Equation~(\ref{eq:totratefin}), yields the
asymptotic high-B result derived by Baier et al. \cite{bms96}, and
reproduced independently by Baring \& Harding \cite{bh97apj}; the
overall rate for \teq{\pll} approaches a value independent of \teq{B}.
Observe that the manifestations of the pair creation threshold for each
of the final photons of \teq{\parallel} polarization (i.e.  at
\teq{\omega'=2} and \teq{\omega''=2}) are the individually-divergent
coefficients of the inverse trigonometric functions.  Yet,
collectively, due to the energy conservation relation \teq{\omega
=\omega'+\omega''}, such divergences cancel each other to yield a
finite overall result as \teq{\omega\to 2}.  For the incident photon of
\teq{\perp} polarization, the pair threshold of \teq{1+\sqrt{1+2 B}} is
remote from \teq{\omega=2} so that it would only become explicitly
apparent when the amplitude was evaluated to higher order in \teq{B}.
Note also that the \teq{\omega\ll 1} limit of
Eq.~(\ref{eq:MperptoparparBgg1}) is \teq{\omega\omega'\omega''/6} and
reproduces results obtained in \cite{Adler71} and \cite{Stone79}.  The
functional form of Eq.~(\ref{eq:MperptoparparBgg1}) is plotted in
Figure~\ref{fig:scattampBgg1}.

The equivalent result for the splitting mode \teq{{\cal M}_{\lpl}}
requires little additional algebra given that it can be obtained from
the analysis just above using the cyclic symmetry transformations of
Eq.~(\ref{eq:permsymm}).  Carefully keeping track of signs and all
photon frequencies by relabelling at the beginning of the
manipulations, the roles of the \teq{\Delta_4} and \teq{\Delta_3} terms
are interchanged, and the obvious result emerges:
\begin{equation}
   {\cal M}_{\lpl}\; \approx \; 
   \dover{4\omega''}{\omega\sqrt{4-\omega^2}}\,\arcsin
      \Bigl(\dover{\omega}{2}\Bigr) -
   \dover{4\omega}{\omega''\sqrt{4-(\omega'')^2}}\,\arcsin 
      \Bigl(\dover{\omega''}{2}\Bigr) + \omega'
   \quad , \quad B\gg 1\;\; .
 \label{eq:MpartoperpparBgg1}
\end{equation}
While not established before in the literature, the low energy limit of
this, namely \teq{{\cal M}_{\lpl}\approx \omega\omega'\omega''/6},
yields the differential rate from previous expositions
\cite{Adler71,Stone79} of low energy approximations.  The form of
Eq.~(\ref{eq:MpartoperpparBgg1}) is displayed in
Figure~\ref{fig:scattampBgg1}, exhibiting the asymmetry expected under
interchanges \teq{\omega'\leftrightarrow\omega''}.  In this case, pair
threshold structure in the amplitude appears again for the two photons
of parallel polarization (i.e. at \teq{\omega=2} and \teq{\omega''=2}),
and is also absent for the produced \teq{\perp} photon, being of higher
order in \teq{B}.  Consequently, the amplitude possesses a real
divergence at \teq{\omega=2}, a noteworthy occurrence that is
illustrated by comparing the two panels of
Figure~\ref{fig:scattampBgg1}.  Such divergences, which are not
integrable over \teq{\omega} (and therefore patently different in
nature from the resonances encountered in rates for \teq{\gamma\to
e^{\pm}}), are characteristic of the photon splitting rate near
resonances at and above the pair threshold of \teq{\omega=2},
corresponding to the creation of virtual pairs in various excited
states.  In fact, near such resonances, photon splitting necessarily
becomes first order in \teq{\fsc} like pair creation as the
intermediate states ``go on-shell.''

\begin{figure}
%
%\vspace{0.2truecm}
\centerline{\psfig{figure=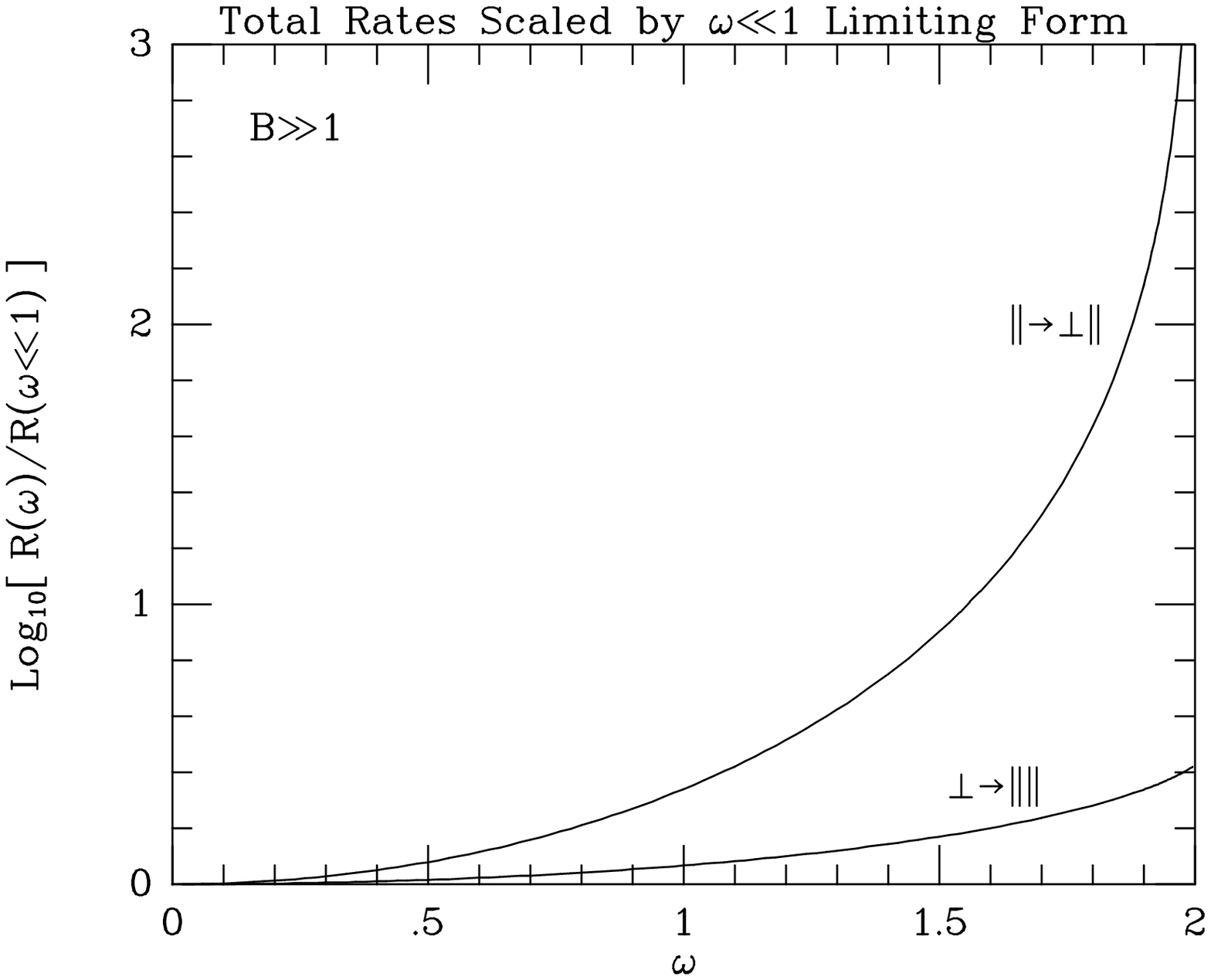,width=12cm}}
\caption{The total rates in the \teq{B\gg 1} limit for the modes
\teq{\perp\to\parallel\,\parallel} and
\teq{\parallel\to\perp\,\parallel}, computed according to
Eq.~(\ref{eq:totratefin}) using the amplitude formulae in
Eqs.~(\ref{eq:MperptoparparBgg1}) and~(\ref{eq:MpartoperpparBgg1}),
divided by the rates that would be computed when taking the low energy
(\teq{\omega\ll 1}) limit of these amplitudes, i.e.  \teq{{\cal
M}_{\perp\to\parallel\,\parallel}\approx \omega\omega'\omega''/6
\approx {\cal M}_{\lpl}}.  Deviations from such low energy
approximations (i.e. \teq{R(\omega)\propto \omega^5}), while
significant for \teq{\perp\to\parallel\,\parallel}, are dramatic for
\teq{\parallel\to\perp\,\parallel} near pair creation threshold
\teq{w=2}.}
 \label{fig:ratesBgg1}
\end{figure}

The rapid increase of the rate of \teq{\lpl} relative to that of
\teq{\pll} is exhibited in Figure~\ref{fig:ratesBgg1}, where the rates
have been scaled by the low energy (\teq{\omega\ll 1}) limiting forms
(\teq{R(\omega )\propto \omega^5}) discussed in the next subsection.
This particular scaling is chosen to illustrate deviations from the
\teq{\omega\ll 1} asymptotic forms, and therefore to demonstrate the
need for relinquishing use of them when sampling photon energies near
pair threshold, a parameter regime very relevant to certain
astrophysical calculations (e.g. see \cite{bh98apjl,hbg97apj}).  The
dominance of the \teq{R_{\lpl}} over \teq{R_{\pll}} near \teq{\omega
=2} apparent in these \teq{B\gg 1} results becomes substantive in
parameter regimes where the weakly-dispersive vacuum (i.e. for
\teq{B\lesssim 1}) polarization selection rules for splitting derived
by Adler \cite{Adler71} (which prohibit \teq{\lpl} and \teq{\ppp}
splittings) may not apply if non-linear contributions to vacuum
polarization or plasma effects are significant.  This underlines the
saliency of a detailed determination of the dispersive properties of
the magnetized vacuum or plasma medium appropriate to a particular
astrophysical scenario.

The derivation of the \teq{B\gg 1} form for the amplitude for
\teq{\ppp} differs significantly from the results just expounded.
First, contributions from \teq{n''=1}, \teq{n=n'=0} and \teq{n'=1},
\teq{n=n''=0} and \teq{n=1}, \teq{n'=n''=0} combinations are
identically equal to zero by virtue of each of the associated
\teq{\Delta_i} factors.  This automatically implies that no inverse
trigonometric functions that have arguments independent of \teq{B}
appear in the amplitude, a property not possessed by the other
splitting modes.  The consequences of this are twofold.  First, this
cancellation implies that the scattering for \teq{\ppp} is of a higher
order in \teq{B} than for the other two splitting modes.  Second, since
any potential appearance of inverse trigonometric functions spawned by
the forms in Eq.~(\ref{eq:Iapprox2}) involves arguments that depend on
\teq{B} through the \teq{{\cal N}}s, these arguments are always small
when \teq{B\gg 1}, precipitating a redundancy with the low energy
limit.  Hence, it follows immediately that the scattering amplitude for
\teq{\ppp} in the regime of highly super-critical fields is identical
to that of the \teq{B\gg 1} specialization of the low energy
(\teq{\omega\ll 1}) limit.  As the latter has been derived in various
papers in the literature (e.g. see \cite{Adler71,Stone79,wbm98} and the
subsequent section), here it is sufficient to merely state the result:
\begin{equation}
   {\cal M}_{\ppp}\; \approx \; 
   \dover{\omega\omega'\omega''}{3B}   \quad , \quad B\gg 1\;\; .
 \label{eq:MperptoperpperpBgg1}
\end{equation}
This extremely simple form differs profoundly from those of the other
two modes because of the absence of photons of \teq{\parallel} 
polarization in the interaction.  Hence any signatures of the pair
threshold of \teq{1+\sqrt{1+2B}} of \teq{\perp} photons are absent
in the domain of \teq{\omega <2}, and a scaling-type form with
obvious cyclic symmetry emerges.

\subsection{Approximations for \teq{\omega\ll 1}}
\label{sec:omegall1}

The low energy limit \teq{\omega\ll 1} is of interest not only because
it was the regime where compact analytic expressions for the splitting
rates were first obtained \cite{Adl70,Bial70,Adler71}, but also because
the analysis that follows derives simple and elegant representations of
the scattering amplitudes in terms of well-known special functions that
provide a convenient alternative option for numerical evaluations.

The amplitudes for each of the splitting modes should exhibit a cubic
energy dependence \cite{Adler71} when \teq{\omega\ll 1}.  Hence, a
necessary product of the Landau representation formalism is that terms
linear in photon energies should contribute exactly zero.  For the
polarization modes \teq{\pll} and \teq{\lpl}, whose amplitudes are
identical in the low energy limit \cite{Adler71,Stone79}, the
demonstration of this is not dissimilar to the \teq{B\gg 1} analysis.
The \teq{\omega\ll 1} restriction generates a single infinite series in
\teq{n} due to the clustering of \teq{n'} and \teq{n''} around
\teq{n}.   The ensuing algebra in the simplification of this series is
moderately lengthy, and requires re-indexing of the logarithmic terms
to assume forms involving \teq{\log_e[1+2nB]}, and also some
relabelling of the rational functions.  Care must be taken in these
rearrangements due to the infinite nature of the series, and the
technique adopted is outlined just below.  The terms linear in photon
energy result in zero, as expected: for more details, the reader is
referred to \cite{wbm98}.  The next order contribution is cubic in
energy, as desired, with terms coming from a mixture of (i) the linear
terms of the \teq{\Delta_i} combined with the quadratic higher order
terms of the \teq{p_z} integrals, and (ii) the cubic \teq{\Delta_i}
terms in conjunction with the leading order (constant) terms from the
\teq{p_z} integrals.  The algebra is straightforward, but lengthy and
tedious, generating an exact cancellation of all but terms proportional
to \teq{\omega\omega'\omega''}.  This approach leads to a reproduction
of the \teq{C_j} listed in entirety in Appendix B of Weise, Baring \&
Melrose \cite{wbm98} for both the \teq{\pll} and \teq{\ppp} modes of
splitting.  Hence there is little point in replicating these
expressions here; the reader is referred to \cite{wbm98} for details.

These results are expressed as single infinite series in the label
\teq{n}, which sometimes starts at \teq{n=0}, and sometimes begins at
higher integer values (up to \teq{3}).  Hence, an aesthetic goal is to
rearrange some of these series so that the summations in each
contribution begin at \teq{n=0}, and then add the terms in the series
together.  This is a non-trivial exercise, given the divergent nature
of the series in many of the individual contributions.  Hence,
considerable care must be taken when performing the rearrangements, for
which there is no unique prescription.  One choice for relabelling the
sums is adopted by Weise, Baring \& Melrose \cite{wbm98}, though their
end results expressed in their Appendix C do not facilitate analytic
development in the most expedient manner, and were in fact erroneous
(discussed briefly below).  An alternative and preferable choice for
rearrangement of the multitude of series over the label \teq{n} is
adopted here, outlined as follows.  Inspection of the various \teq{C_j}
contributions in Appendix B of \cite{wbm98} reveals that they always
consist of three types of terms: (i) logarithmic ones proportional to
\teq{\log_e[1+2(n+l)B]}, for \teq{l=0,\pm 1,\pm 2,\pm 3}, (ii) rational
functions of \teq{1+2(n+l)B} for \teq{l=0,\pm 1,\pm 2}, and (iii)
polynomials in \teq{n}.  A unique method for rearrangement is to
truncate all series to finite ones with \teq{n\leq k}, and then perform
relabellings so that the first two types of terms consist only of
\teq{\log_e[1+2nB]} terms and rational functions of \teq{1+2nB}.  This
approach provides no particular focus on series that originate with
labels \teq{n>0}, but requires careful accounting of the remainder
terms at the upper and lower ends of the sums, for which significant
cancellation arises.  The coefficients of the logarithmic functions,
originally cubic in \teq{n}, reduce to linear functions of \teq{n} in
this development.  The consequent simplification of the series terms is
counterbalanced by the transferral of complexity to the constant
remainder terms, which are purely functions of \teq{k} and \teq{B}.
Taking the limit of \teq{k\to\infty} achieves the desired (and
convergent) result.

After considerable algebra collecting together all the constituent
series in Appendix B of \cite{wbm98}, and performing the rearrangement
as just prescribed, one arrives at the following series representation
of the scaled scattering amplitudes:
\begin{equation}
   {\cal M}_{e\to e'e''}\; =\;
   \omega\omega'\omega''\; \lim_{k\to\infty}\; \Biggl\{\,
   \sum_{n=0}^{k} T_{e\to e'e''}(n,\, B)
      + {\cal R}_{e\to e'e''}(k,\, B)\; \Biggr\}
 \label{eq:Mseries1}
\end{equation}
for \teq{\omega\ll 1}, where
\begin{eqnarray}
   T_{\pll}(n,\, B) & = &
   - \biggl(\dover{4n}{B} + \dover{3}{2B^2}\biggr)
                          \log_e\Bigl[n+\dover{1}{2B}\Bigr]
   + \biggl( \dover{2}{3}+\dover{1}{2B^2} \biggr)\; \dover{1}{1+2nB}
   - \dover{1}{3(1+2nB)^2}\nonumber\\[-5.5pt]
 \label{eq:Mseries1Tdef}\\[-5.5pt]
   T_{\ppp}(n,\, B) & = &
   \dover{3}{2B^2} \log_e\Bigl[n+\dover{1}{2B}\Bigr]
   - \dover{3}{2B^2}\; \dover{1}{1+2nB}
   + \dover{1}{2B^2} \; \dover{1}{(1+2nB)^2}\nonumber
\end{eqnarray}
defines the series terms.  The remainders are quite lengthy, and are
listed in Appendix A.

Consider first the polarization mode \teq{\pll}.  While possibly only
marginally simpler than Eq.~(C1) of \cite{wbm98}, the series and
remainder in Eqs.~(\ref{eq:Mseries1}), (\ref{eq:Mseries1Tdef})
and~(\ref{eq:Mseries1Rdef}) naturally enable the development of a
special function representation of the scattering amplitude.  The
finite summation over terms like \teq{(x+n) \log_e(x+n)} in
Eq.~(\ref{eq:Mseries1Tdef}) can be expressed using result 44.1.2 of
\cite{hans75} in terms of an integral of the logarithm of the Gamma
function.  At this juncture, the analysis begins to image parts of that
generated in expressing the polarization properties of a magnetized
vacuum via effective Lagrangian or proper-time techniques
\cite{te75,dtz79,ivan92}, as should be expected.  Hence, it is
appropriate to adopt definitions from such literature as much as
possible.  Following \cite{te75,dtz79}, here a definition for the
generalized Gamma function \teq{\Gamma_1(x)} of
\begin{equation}
  \log_e\Gamma_1(x)\; =\; \int_0^x dt\,\log_e\Gamma (t)
  + \dover{1}{2}\, x(x-1) - \dover{x}{2}\log_e2\pi
 \label{eq:Gamma1def}
\end{equation}
is adopted.  Properties of this function, which include
\teq{\Gamma_1(1)=1}, are discussed at length in \cite{bend33}
and outlined in Appendix B.  

Using Eq.~(\ref{eq:sumnlogn}), one soon arrives at an expression for
the scattering amplitude in terms of a handful of special (polygamma)
functions, namely \teq{\Gamma_1(x)}, and \teq{\log_e\Gamma (x)} and its
derivatives.  This representation consists of two parts, one
independent of \teq{k}, and one that involves a limit as
\teq{k\to\infty} of the remainder in Eq.~(\ref{eq:Mseries1Rdef}),
combined with several terms incorporating the special functions with
arguments that depend on \teq{k}.  In evaluating this limit, most terms
can be handled in a straightforward manner, and standard asymptotic
series (e.g. see \cite{gr80}) for \teq{\log_e\Gamma (x)} and \teq{\psi
(x)} as \teq{x\to\infty} prove useful.  However, the treatment of the
term involving the function \teq{\log_e\Gamma_1(1+k+1/2B)} that appears
in the limit contribution is non-trivial.  A series representation for
this function for large arguments is required, and is presented in
Eq.~(\ref{eq:logGamma1asymp}).  Assembling the various pieces, the
limiting result as \teq{k\to\infty} is
\begin{eqnarray}
 {\cal M}_{\pll} & = & \omega\omega'\omega''\;
   \Biggl\{\, \dover{4}{B} \log_e\Gamma_1\Bigl(\dover{1}{2B}\Bigr)
      - \dover{1}{2B^2} \log_e\Gamma\Bigl(\dover{1}{2B}\Bigr)
      - \biggl( \dover{1}{3B} + \dover{1}{4B^3}\biggr)
                     \psi \Bigl(\dover{1}{2B}\Bigr)  \nonumber\\[-6pt]
 \label{eq:Mperptoparparwll1}\\[-6pt]
   && \quad 
   - \dover{1}{12B^2} \,\psi' \Bigl(\dover{1}{2B}\Bigr) 
      - \dover{1}{6} + \dover{1}{6B} - \dover{4 L_1}{B}
      + \dover{1}{4B^2} \Bigl( \log_e2\pi -1 -3 \log_e 2B \Bigr) 
        \Biggr\}   \nonumber
\end{eqnarray}
for \teq{\omega\ll 1}.  This is the sought-after compact analytic form
that is comparable in simplicity to the one-loop effective Lagrangians
calculated in \cite{te75,dtz79}.  Using series and asymptotic
expansions for all the special functions present, it is routine to
establish that \teq{{\cal M}_{\pll}\approx (26
B^3/315)\,\omega\omega'\omega''} for \teq{B\ll 1}, while for \teq{B\gg
1}, one finds \teq{{\cal M}_{\pll}\approx
\omega\omega'\omega''/6}, a result obtainable from
Eq.~(\ref{eq:MperptoparparBgg1}).

The developments are similar for the \teq{\ppp} mode:  this
representation again consists of two parts, one independent of \teq{k}, and
one that involves a limit as \teq{k\to\infty} of the remainder in
Eq.~(\ref{eq:Mseries2Rdef}), combined with several terms incorporating
polygamma functions with arguments that depend on \teq{k}.  This limit
can easily be evaluated using asymptotic series to yield
(for \teq{\omega\ll 1})
\begin{eqnarray}
 {\cal M}_{\ppp} & = & \omega\omega'\omega''\;
   \Biggl\{\, - \dover{3}{2B^2} \log_e\Gamma\Bigl(\dover{1}{2B}\Bigr)
      + \dover{3}{4B^3} \psi \Bigl(\dover{1}{2B}\Bigr)  \nonumber\\[-6pt]
 \label{eq:Mperptoperpperpwll1}\\[-6pt]
   && \quad 
   + \dover{1}{8B^4} \,\psi' \Bigl(\dover{1}{2B}\Bigr) 
      + \dover{1}{3B} + \dover{1}{2B^2} - \dover{1}{B^3}
      + \dover{3}{4B^2} \Bigl( \log_e2\pi + \log_e 2B \Bigr) 
        \Biggr\}\;\; .   \nonumber
\end{eqnarray}
Using series and asymptotic expansions for all the special functions
present, it is routine to establish that \teq{{\cal M}_{\ppp}\approx
(48 B^3/315)\,\omega\omega'\omega''} for \teq{B\ll 1}, while for
\teq{B\gg 1}, one finds \teq{{\cal M}_{\ppp}\approx
\omega\omega'\omega''/(3B)}, the result stated in
Eq.~(\ref{eq:MperptoperpperpBgg1}).

It must be remarked in passing that the expressions for \teq{\ppp} in
Eqs.~(\ref{eq:Mseries1Tdef}), (\ref{eq:Mseries2Rdef})
and~(\ref{eq:Mperptoperpperpwll1}) cannot be derived from the series in
Eq.~(C2) of \cite{wbm98}, principally because that series expression is
divergent, and therefore erroneous.  Such an error was introduced by an
inappropriate rearrangement of individually-divergent contributing
series (leading to the addition of infinite contributions), a mistake
that is avoided by the careful technique employed here in manipulating
the results of Appendix B in \cite{wbm98}.  Notwithstanding, the
numerical results for the \teq{\ppp} mode presented in
\cite{wbm98} were effectively evaluated before any series
rearrangement, and therefore remain valid.

\begin{figure}
\vspace{0.2truecm}
\centerline{\psfig{figure=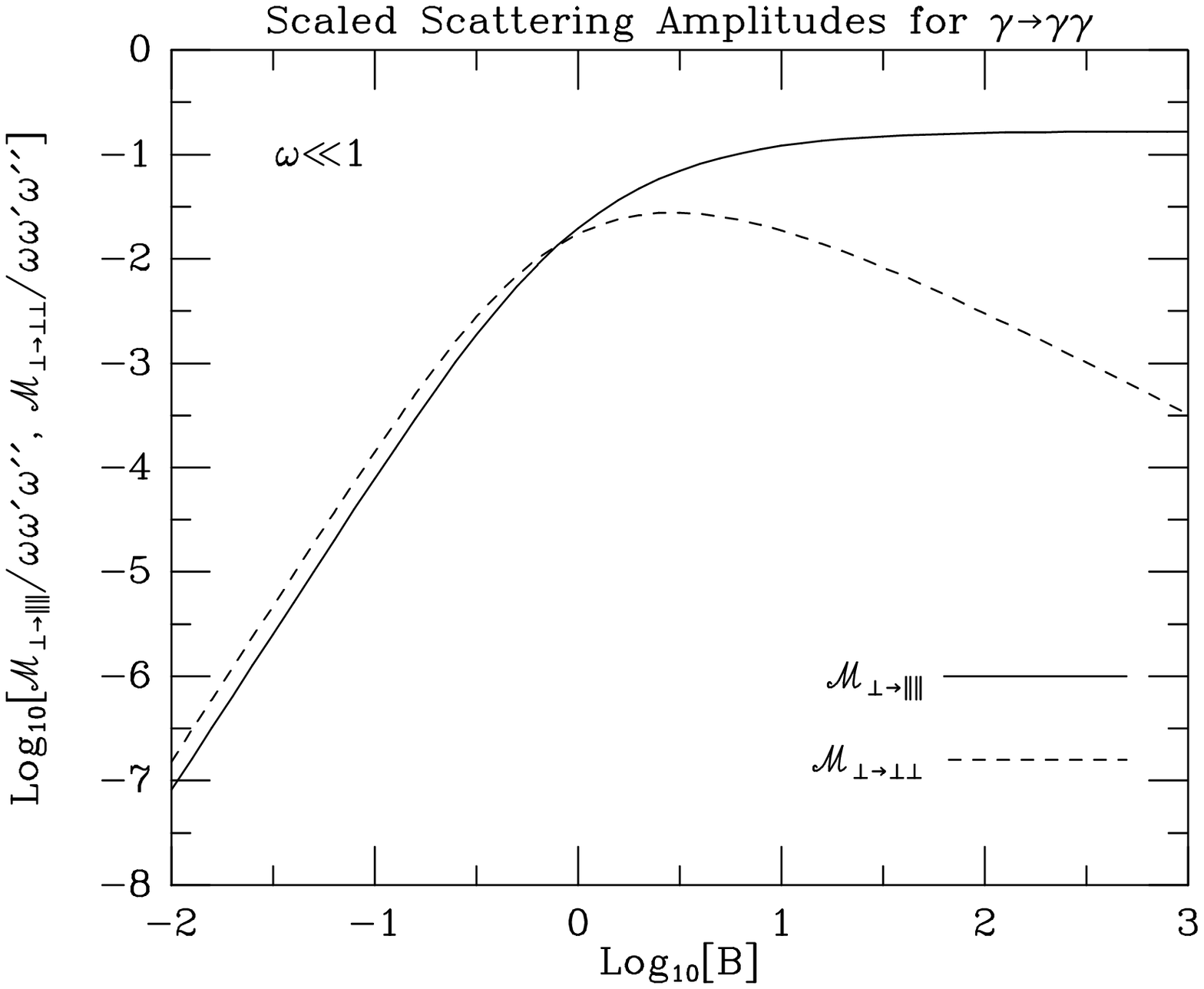,width=11cm}}
\caption{The dependence of the scattering amplitudes, scaled by
\teq{\omega\omega'\omega''}, on the magnetic field (in units of
\teq{B_{\rm c}=4.413\times 10^{13}}Gauss) for the splitting modes
\teq{\perp\rightarrow\parallel\,\parallel} (see
Eq.~[\ref{eq:Mperptoparparwll1}]) and
\teq{\perp\rightarrow\perp\,\perp} (see
Eq.~[(\ref{eq:Mperptoperpperpwll1}]), for photon energies well below
pair creation threshold.  The amplitude for
\teq{\parallel\rightarrow\perp\,\parallel} is identical to that for
\teq{\perp\rightarrow\parallel\,\parallel}.}
 \label{fig:scattampwll1}
\end{figure}

The compact analytic forms presented in
Eqs.~(\ref{eq:Mperptoparparwll1}) and~(\ref{eq:Mperptoperpperpwll1})
represent the culmination of the \teq{\omega\ll 1} focus here.  It is
the first time such simple forms for the scattering amplitudes
involving just special functions have been calculated in this limit,
though somewhat more convoluted, yet essentially equivalent,
expressions have been put forward in \cite{hh97}.  A distinct advantage
of the expressions in Eqs.~(\ref{eq:Mperptoparparwll1})
and~(\ref{eq:Mperptoperpperpwll1}) is the ease with which they can be
accurately computed numerically.  Their dependence on \teq{B} is
illustrated in Figure~\ref{fig:scattampwll1}, replicating the numerics
of \cite{wbm98} and earlier effective Lagrangian determinations
\cite{Adler71}, which are just as expedient since (see
Eqs.~[\ref{eq:Mperptoparparwll1_ELP}]
and~[\ref{eq:Mperptoperpperpwll1_ELP}] below) they involve just
integrals of elementary functions.

The low frequency result for \teq{{\cal M}_{\parallel\to\perp
\parallel}} is not presented explicitly since it reproduces that for
\teq{{\cal M}_{\pll}} (e.g.  see \cite{Adler71,Stone79}); this is due
to the crossing symmetries involved.  Note that while the cubic
dependences of all the modes at low energies reflect the lack of an
energy scale in this domain (i.e.  such as pair threshold), the
normalizations are dependent on the polarization mode, particularly at
highly supercritical fields where the \teq{{\cal M}_{\ppp}} amplitude
is highly suppressed.  This effectively represents how the rate
normalization is sensitive to the (virtual) pair creation thresholds
for the polarization states involved in a particular splitting mode.

To conclude this presentation focusing on the \teq{\omega\ll 1}
specialization, an obvious objective is the re-derivation of
Eqs.~(\ref{eq:Mperptoparparwll1}) and (\ref{eq:Mperptoperpperpwll1})
starting with extant and well-known effective Lagrangian/proper-time
(ELP) results, and thereby demonstrating analytically the equivalence
of the S-matrix formulation in the Landau representation and
Schwinger-type formalisms in the low energy limit.  Consider first the
mode \teq{\pll}, for which such a determination is somewhat involved.
The starting point is the integral expression
\cite{Adler71,Stone79,bh97apj} that corresponds to the scaled
scattering amplitude that generates the same form for the rate as in
Eq.~(\ref{eq:totratefin}):
\begin{equation}
   {\cal M}_{\pll}^{\hbox{\sevenrm ELP}} \; =\;
     \dover{\omega\omega'\omega''}{B}\;
     \int^{\infty}_{0} \dover{ds}{s}\, e^{-s/B}\,
   \Biggl\{ \biggl(-\dover{3}{4s}+\dover{s}{6}\biggr)\,\dover{\cosh s}{\sinh s} 
   +\dover{3+2s^2}{12\sinh^2s}+\dover{s\cosh s}{2\sinh^3s}\Biggr\}\; , 
 \label{eq:Mperptoparparwll1_ELP}
\end{equation}
which has \teq{B\ll 1} and \teq{B\gg 1} limits matching those of
Eq.~(\ref{eq:Mperptoparparwll1}).  In the subsequent analysis, it is
useful to manipulate integrations using the variable \teq{\mu =1/B}.
The first step is to recognize that \teq{1/s^3} times the factor
in curly braces in Eq.~(\ref{eq:Mperptoparparwll1_ELP}) is a perfect
derivative, namely \teq{dg/ds}, where
\teq{g(s)= (1/4s^3)\; d[s \coth s-1]/ds - \coth s/(6s)}.
Integration by parts is obviously the operative method, with the goal
of retaining \teq{\coth s} functions explicitly, combined with 
powers of \teq{s}.  After some algebra, one finds that
\begin{eqnarray}
   \dover{{\cal M}_{\pll}^{\hbox{\sevenrm ELP}}}{
     \omega\omega'\omega''} & = & 
    \dover{1}{B} \int^{\infty}_{0} ds\, e^{-s/B}\,
     \biggl[ \dover{1}{4B^2s} - \dover{1}{4Bs^2} - \dover{1}{6B}
         + \dover{1}{3s} \biggr]\; (s \coth s-1) \nonumber\\[-6pt]
 \label{eq:Mperptoparparwll1_ELP2}\\[-6pt]
   && \qquad - \dover{1}{B} \int^{\infty}_{0} \dover{ds}{s^3}\, e^{-s/B}\,
     \biggl[s \coth s-1 - \dover{s^2}{3} \biggr]  \nonumber
\end{eqnarray}
results.  The integral on the first line can be performed using
identities 3.551.3 and 3.554.4 of \cite{gr80}, yielding the Gamma
function and polygamma functions (or equivalently generalized Riemann
Zeta functions) in addition to elementary functions.  The only subtle
part pertains to the second term of this integral, namely that
contributed by the \teq{-1/(4Bs^2)} factor.  This can be differentiated
with respect to \teq{B}, evaluated to yield a \teq{\psi} function, and
then the result integrated, noting the behaviour as \teq{B\to 0}.  The
evaluation of the integral on the second line of
Eq.~(\ref{eq:Mperptoparparwll1_ELP2}) is much more involved.  However,
it has been performed before in the literature, and appears explicitly
in calculations \cite{te75,dtz79} of the one-loop effective Lagrangian
describing refractive indices of the magnetized vacuum in QED.  Hence
the motivation for the particular partitioning of integrations chosen in
Eq.~(\ref{eq:Mperptoparparwll1_ELP2}).  Details of the determination of
this integral are found in Dittrich et al. \cite{dtz79}, and the second
line of Eq.~(\ref{eq:Mperptoparparwll1_ELP2}) can be equated to
\teq{-8\pi^2/B^3} times the Lagrangian \teq{{\cal L}^{(1)}(B)} (see
Eqs.~(2.4) and (3.16) of \cite{dtz79}), thereby introducing the
\teq{\Gamma_1} function.  Collecting together the terms neatly
generates an analytic form for \teq{{\cal
M}_{\pll}^{\hbox{\sevenrm ELP}}} that is
identical to Eq.~(\ref{eq:Mperptoparparwll1}), so that the desired
demonstration of equivalence of the Landau representation and effective
Lagrangian forms is achieved.

The procedure for the \teq{\ppp} mode is similar, though somewhat less
involved.  The equivalent scaled scattering amplitude obtained
\cite{Adler71,Stone79} from effective Lagrangian/proper-time techniques
is
\begin{equation}
   {\cal M}_{\ppp}^{\hbox{\sevenrm ELP}} \; =\;
     \dover{\omega\omega'\omega''}{B}\;
     \int^{\infty}_{0} \dover{ds}{s}\, e^{-s/B}\,
   \Biggl\{ \dover{3}{4s}\,\dover{\cosh s}{\sinh s} 
   +\dover{3-4s^2}{4\sinh^2s}-\dover{3s^2}{2\sinh^3s}\Biggr\}\; . 
 \label{eq:Mperptoperpperpwll1_ELP}
\end{equation}
Recognizing that the factor in curly braces can be written as
\teq{-(3s/4)\; d[\coth s/s-1/s^2]/ds + (s^2/4)\; d^3[\coth
s-1/s]/ds^3}, integration by parts is again indicated, with identities
3.551.3 and 3.554.4 of \cite{gr80} again proving useful.  With
manipulations similar to (but simpler than: the \teq{\Gamma_1} function
is not involved here) those for the \teq{\pll}, a modicum of algebra
leads to the derivation of Eq.~(\ref{eq:Mperptoperpperpwll1}) from
Eq.~(\ref{eq:Mperptoperpperpwll1_ELP}), as desired.  This equivalence
is a satisfying indication of the verity of the Landau representation
analysis in this paper.

\section{Conclusion}
\label{sec:conclusion}

This paper has provided a detailed development of the S-matrix
formulation of the QED process of magnetic photon splitting in the
Landau representation, focusing on the case of zero dispersion where
photon propagation is collinear.  The formalism in
Section~\ref{sec:formalism} rederives and extends the exposition of
Mentzel, Berg \& Wunner \cite{mbw94}.  The two principal general
developments offered here are an analytic reduction via the summation
over the spins of the intermediate pair states, discussed briefly in
\cite{wbm98}, and the analytic integration over the momenta parallel to
the field incorporated in the electron propagators.  This latter
accomplishment is presented here for the first time.  The cumulative
product of these developments is a satisfyingly simple and elegant form
in Eq.~(\ref{eq:Mform}) for the scattering amplitude for each of the
polarization modes permitted by CP invariance.  These amplitudes
possess products of generalized Laguerre polynomials that are common to
QED processes in external magnetic fields, and elementary functions
involving the photon energies and the various pair thresholds
associated with the propagators.  Moreover, the analytic forms
presented consist of just triple summations over Landau level quantum
numbers of the intermediate states, and are eminently suitable for
accurate numerical computations both below and above pair creation
threshold \teq{\omega =2}.  The applicability of these results to
regimes above pair threshold is a benefit of the S-matrix expansion in
the Landau representation that is not afforded by effective Lagrangian
and proper-time calculations:  while these (latter) Schwinger-type
techniques elegantly formulate splitting rates below pair threshold,
they eliminate the resonance structure early on in their mathematical
developments, a severe limitation above \teq{\omega =2}.

As an embellishment to these general results, specializations in two
significant domains have been obtained.  The first is for highly
supercritical fields, \teq{B\gg 1}, reproducing in particular the
result of \cite{bms96} for the \teq{\pll} mode, and deriving new
results for the other two modes permitted by CP invariance in the limit
of zero dispersion.  The second group of asymptotic results are for
energies \teq{\omega\ll 1} well below pair creation threshold, where
new and compact expressions for the scattering amplitudes have been
derived in Eqs.~(\ref{eq:Mperptoparparwll1})
and~(\ref{eq:Mperptoperpperpwll1}) in terms of the logarithm of the
Gamma function, its integral and their derivatives.  These two domains
of specialization herein have facilitated the first analytic
demonstration of the equivalence of splitting rates obtained by the
S-matrix formulation in the Landau representation and those derived
using Schwinger-type effective Lagrangian/proper-time techniques.

\acknowledgements
I thank my collaborators Jeanette Weise, Don Melrose and Alice Harding,
and also Carlo Graziani and George Pavlov for many productive
discussions, and also the referee Stephen Adler for suggestions
that improved the presentation.  I also thank the RCfTA at the
University of Sydney for hospitality during part of the period when
this work was performed.

\appendix

\section{}  %   Appendix A           AAAAAAAAAAAAAAAAAAA

Here the remainders that appear in the series representation
in Eq.~(\ref{eq:Mseries1}) for the \teq{\omega\ll 1} specializations
to the splitting amplitudes for polarization modes \teq{\pll} and
\teq{\ppp} are presented:
\begin{eqnarray}
   {\cal R}_{\pll}(k,\, B) & = &
   \dover{k+1}{8B} \biggl[ 2k^2 
           + k \Bigl( 4 + \dover{1}{B}\Bigr)
           + \dover{1}{B}\biggr] \log_e\Bigl[k+2+\dover{1}{2B}\Bigr]
   \nonumber\\
    & + & \dover{1}{4B} \biggl[ 11k^3 + k^2 \Bigl( 21 + \dover{21}{2B}\Bigr)
         + k \Bigl( 10 + \dover{14}{B} + \dover{5}{2B^2}\Bigr)
           + \dover{5}{2B} + \dover{5}{4B^2} \biggr] 
                           \log_e\Bigl[k+1+\dover{1}{2B}\Bigr]
   \nonumber\\
    & - & \dover{1}{4B} \biggl[ 11k^3 + k^2 \Bigl( 16 + \dover{21}{2B}\Bigr)
           + k \Bigl( 5 + \dover{9}{B} + \dover{5}{2B^2}\Bigr)
           + \dover{5}{4B^2} \biggr] 
                           \log_e\Bigl[k+\dover{1}{2B}\Bigr]\nonumber\\[-5.5pt]
 \label{eq:Mseries1Rdef}\\[-5.5pt]
    & - & \dover{k}{4B} \biggl[ k^2 + \dover{k}{2B} - 1 \biggr]
                           \log_e\Bigl[k-1+\dover{1}{2B}\Bigr]
   \nonumber\\
    & - & \biggl( \dover{B}{3} (k+1) + \dover{1}{6} + \dover{k+1}{2B} \biggl)\;
             \dover{1}{1+2(k+1)B} + \dover{k}{2B[1+2kB]}
   \nonumber\\
    & - & \dover{9k^2}{2B} - \dover{5k}{B} - \dover{7k}{4B^2} - \dover{5}{8B}
    - \dover{9}{8B^2} - \dover{3}{4B^2}\,\log_e 2B\;\; ,
   \nonumber
\end{eqnarray}
and
\begin{eqnarray}
   {\cal R}_{\ppp}(k,\, B) & = &
   - \dover{(k+1)(k+2)(k+3)}{12B}\; \log_e\Bigl[k+3+\dover{1}{2B}\Bigr]
   \nonumber\\
   && -\dover{k+1}{16B} \biggl[ \dover{13}{3} k^2
           + k \Bigl( \dover{32}{3} + \dover{7}{2B}\Bigr)
           + 4 + \dover{7}{2B}\biggr] \log_e\Bigl[k+2+\dover{1}{2B}\Bigr]
   \nonumber\\
   && +\dover{1}{16B} \biggl[ \dover{71}{3} k^3
           + k^2 \Bigl( 38 + \dover{9}{2B}\Bigr)
           + k \Bigl( \dover{79}{3} - \dover{4}{B}\Bigr)
           + 12 - \dover{2}{B}\biggr] \log_e\Bigl[k+1+\dover{1}{2B}\Bigr]
   \nonumber\\
   && -\dover{1}{16B} \biggl[ \dover{71}{3} k^3
           + k^2 \Bigl( 13 + \dover{9}{2B}\Bigr)
           - k \Bigl( \dover{20}{3} - \dover{13}{B}\Bigr)
           + \dover{13}{2B}\biggr] \log_e\Bigl[k+\dover{1}{2B}\Bigr]
   \nonumber\\[-6pt]
 \label{eq:Mseries2Rdef}\\[-6pt]
   && +\dover{k}{16B} \biggl[ \dover{13}{3} k^2 
           + k \Bigl( 2 + \dover{7}{2B}\Bigr)
           - \dover{19}{3}\biggr] \log_e\Bigl[k-1+\dover{1}{2B}\Bigr]
   \nonumber\\
  &&  \;\; +\dover{k(k-1)(k-2)}{12B}\; \log_e\Bigl[k-2+\dover{1}{2B}\Bigr]
           + \dover{(k+1)(k+2)^2}{2[1+2(k+2)B]}
   \nonumber\\
  &&  \;\; - \dover{(k+1)^2(5k+2)}{2[1+2(k+1)B]}
           - \dover{k^2(5k-1)}{2[1+2k B]}
           + \dover{k(k-1)^2}{2[1+2(k-1)B]}
   \nonumber\\
 && \quad + \dover{7k^2}{4B} + \dover{7k}{4B} + \dover{9}{8B}
          + \dover{k+1}{B^2} + \dover{3}{4B^2}\,\log_e2B\;\; .
   \nonumber
\end{eqnarray}

\section{}  %   Appendix B           BBBBBBBBBBBBBBBBBBB

In this Appendix, various useful properties of the \teq{\Gamma_1(x)}
function, the integral of the logarithm of the Gamma function, that are
needed in the \teq{\omega\ll 1} specializations are stated.  Given the
definition of \teq{\Gamma_1} in Eq.~(\ref{eq:Gamma1def}), it is
elementary to establish, using 44.1.2 of \cite{hans75}, that
\begin{equation}
   \sum_{n=0}^k (x+n)\log_e(x+n) \; =\;
   \log_e\Gamma_1(1+x+k) - \log_e\Gamma_1(x)\quad .
 \label{eq:sumnlogn}
\end{equation}
Taking successive derivatives with respect to \teq{x}, one quickly arrives
at well-known finite series representations of \teq{\Gamma (x)}
and its logarithmic derivative \teq{\psi (x)}; see \cite{gr80,erd81}
for discussions of these functions and their series representations.

An asymptotic series representation for the \teq{\Gamma_1} function for large
arguments is useful, and can be derived with the aid of the following
series representation (see result 8.343.2 of \cite{gr80}) for the
logarithm of the Gamma function:
\begin{equation}
   \log_e\Gamma (x) \; =\; \Bigl( x-\dover{1}{2} \Bigr)\, \log_ex - x 
      + \dover{1}{2}\log_e 2\pi 
    + \dover{1}{2} \sum_{m=1}^{\infty} \dover{m}{(m+1)(m+2)}\;
       \biggl[ \zeta (m+1,x) - \dover{1}{x^{m+1}} \biggr]
 \label{eq:logGammaseries}
\end{equation}
from which Stirling's asymptotic expansion can be derived.  Here,
\teq{\zeta (m,t)} is the generalized Riemann Zeta function, defined in
9.511 and 9.521.1 of \cite{gr80}.  The integration of this series is
effected using the identity \teq{\zeta'(m,t) = -\zeta (m+1,t)}, and is
mostly uneventful.  However, the treatment of the \teq{m=1} term in the
summation is somewhat more subtle, due to the singular nature of
\teq{\zeta (0,t)}, and requires taking the limit \teq{m\to 1^+},
assuming \teq{m} to be a continuous variable.  Then result 8.362.1 of
\cite{gr80} comes in handy, and the series identity
\begin{eqnarray}
   \log_e\Gamma_1(x) & = & \dover{x-1}{4}\; \Bigl( 2x\,\log_ex - x -1 \Bigr) 
      + \dover{1}{12} \, \Bigl[ \psi (x) + \dover{1}{x}
      + \gamma_{\hbox{\sevenrm E}} - 1 \Bigl]  \nonumber\\[-6pt]
 \label{eq:logGamma1series}\\[-6pt]
    && \quad + \dover{1}{2} \sum_{m=2}^{\infty} \dover{1}{(m+1)(m+2)}\;
       \biggl[ \zeta (m,1) - \zeta (m,x) - 1 + \dover{1}{x^m} \biggr]  
    \nonumber
\end{eqnarray}
follows, where \teq{\gamma_{\hbox{\sevenrm E}}=-\psi (1)\approx 0.5772}
is Euler's constant.  This series, which adequately substitutes for an
asymptotic representation, can be used very effectively for numerical
evaluations for all \teq{x\geq 1}.  For the range \teq{0\leq x < 1},
this series also effects accurate evaluation of \teq{\log_e\Gamma_1(x)}
via use of the recurrence relation
\teq{\log_e\Gamma_1(x)=\log_e\Gamma_1(1+x) -x\log_e x}, an identity
derivable from Eq.~(\ref{eq:Gamma1def}) with the aid of 6.441.3 in
\cite{gr80}. For large \teq{x}, it then follows that
\begin{equation}
   \log_e\Gamma_1(x) \;\sim\;
   \biggl( \dover{x(x-1)}{2} + \dover{1}{12} \biggr) \,\log_ex
     - \dover{x^2}{4} + L_1 + O\Bigl(\dover{1}{x}\Bigr)
 \label{eq:logGamma1asymp}
\end{equation}
where
\begin{eqnarray}
   L_1 & = & \dover{\gamma_{\hbox{\sevenrm E}}}{12} + \dover{1}{6}
           + \dover{1}{2} \sum_{m=2}^{\infty}
                \dover{\zeta (m,1) - 1}{(m+1)(m+2)}   \nonumber\\[-6pt]
 \label{eq:L1def}\\[-6pt]
       & \equiv & \lim_{k\to\infty} \Biggl\{
           \sum_{n=1}^k n\,\log_e n - \biggl( \dover{k(k+1)}{2}
              + \dover{1}{12} \biggr) \, \log_e k + \dover{k^2}{4} \Biggr\}
   \nonumber
\end{eqnarray}
with numerical value \teq{L_1\approx 0.24875}.  This is just the
constant appearing in the magnetized vacuum polarization analyses of
\cite{te75,dtz79}, where the Raabe integral form for it can be found.
The second definition of \teq{L_1} in Eq.~(\ref{eq:L1def}) can be
obtained by setting \teq{x=0} in Eq.~(\ref{eq:sumnlogn}), and is a
result noted by \cite{hh97}.

\end{document}